\documentclass[12pt,onecolumn, draftclsnofoot]{IEEEtran}
\usepackage{cite}
\usepackage{graphicx}
\usepackage{psfrag}
\usepackage{stfloats}
\usepackage[config, textfont=small]{caption,subfig}
\usepackage{color}
\usepackage[english]{babel}
\usepackage{url}
\usepackage{dsfont}
\usepackage{enumitem}
\usepackage{amsmath}
\usepackage{amssymb}
\usepackage{array}
\usepackage{graphics}
\usepackage{epsfig}
\usepackage{amsbsy}
\usepackage{amssymb}
\usepackage{amsthm}
\usepackage{framed}
\usepackage{algorithm}

\allowdisplaybreaks

\newtheorem{theorem}{Theorem}
\newtheorem{lemma}{Lemma}

\newtheorem{remark}{Remark}
\newtheorem{definition}{Definition}
\newtheorem{corollary}{Corollary}
\newtheorem{conjecture}{Conjecture}
\def\smallint{\begingroup\textstyle \int\endgroup}

\usepackage{authblk}

\begin{document}
\title{The Transdimensional Poisson Process  \\ for  Vehicular Network Analysis}
\author{\IEEEauthorblockN{Jeya Pradha Jeyaraj, \textit{Graduate Student Member, IEEE,} Martin Haenggi, \textit{Fellow, IEEE}, Ahmed Hamdi Sakr, \textit{Senior Member, IEEE}, and Hongsheng Lu} \\
\thanks{J. P. Jeyaraj and M. Haenggi are with the Department of Electrical Engineering, University of Notre Dame, Notre Dame, IN 46556, USA. E-mail: $\lbrace$jjeyaraj, mhaenggi$\rbrace$@nd.edu.

A. H. Sakr is with the Department of Electrical and Computer Engineering, 
University of Windsor, Windsor, ON N9B 3P4, Canada. This work was done when A. H. Sakr was with Toyota Motor North America Research and Development - InfoTech Labs, Mountain View, CA 94043, USA. E-mail: ahmed.sakr@uwindsor.ca.

H. Lu is with Toyota Motor North America Research and Development - InfoTech Labs, Mountain View, CA 94043, USA. E-mail: hongsheng.lu@toyota.com. 

This work was supported in part by Toyota Motor North America and by the U.S.~National Science Foundation (grant CCF 2007498).

Part of this work was presented at the 2019 IEEE Global Communications Conference (GLOBECOM'19)~\cite{jeyaj_tppp}.}}
\maketitle

\begin{abstract}
A comprehensive vehicular network analysis requires modeling the street system and vehicle locations. Even when Poisson point processes (PPPs) are used to model the vehicle locations on each street, the analysis is barely tractable. That holds for even a simple average-based performance metric---the success probability, which is a special case of the fine-grained metric, the meta distribution (MD) of the signal-to-interference ratio (SIR). To address this issue, we propose the transdimensional approach as an alternative. Here, the union of 1D PPPs on the streets is simplified to the transdimensional PPP (TPPP), a superposition of 1D and 2D PPPs. The TPPP includes the 1D PPPs on the streets passing through the receiving vehicle and models the remaining vehicles as a 2D PPP ignoring their street geometry. Through the SIR MD analysis, we show that the TPPP provides good approximations to the more cumbrous models with streets characterized by Poisson line/stick processes; and we prove that the accuracy of the TPPP further improves under shadowing. Lastly, we use the MD results to control network congestion by adjusting the transmit rate while maintaining a target fraction of reliable links. A key insight is that the success probability is an inadequate measure of congestion as it does not capture the reliabilities of the individual links.
\end{abstract}


\section{Introduction}
\subsection{Background}
The key objective of vehicular networks is to increase traffic safety by enabling vehicles and infrastructure nodes to broadcast safety messages. The messages contain individual vehicle-related information such as a vehicle's speed, position, orientation, etc., and general traffic-related information on danger zones, weather hazards, etc. The success probability or packet reception rate is the commonly used metric to analyze the performance of vehicular communication. It is defined as the probability that a vehicle can successfully receive a packet from a broadcasting entity at a certain distance. Stochastic geometry~\cite{haenggi} provides the mathematical tools to model vehicular networks by characterizing different street patterns and uncertainties in the vehicle locations on the streets and analyze the performance metrics of interest.

There exist many possibilities to model streets for different geographical regions. A prominent line-based street model is the Poisson line process (PLP)~\cite{baccelli, choi,stoyan, dhillon_book}, which represents the street system as a countable collection of lines. PLPs allow modeling streets with different orientations, making them relevant in characterizing regular and irregular street patterns~\cite{gloag}. For example, a part of the Manhattan city can be modeled by setting the orientations of the lines to $\lbrace 0, \pi/2\rbrace$. 
For a city that has streets of different lengths, line segments/sticks can better characterize the streets than infinitely long lines. One such stick-based model is the Poisson stick process (PSP)~\cite{jeyaj_equiv}, which is a countable collection of sticks with the lengths and orientations of the sticks following some distributions. Another notable stick-based model is the Poisson lilypond model (PLM). In the PLM, the sticks grow at a constant rate until they hit another stick like lilies in a pond. Such a touch-and-stop growth mechanism forms T-junctions, in contrast to the PLP and PSP, which form intersections. The PSP is a versatile model whose parameters can be modified to either accurately or approximately characterize the PLP, PLM, and their rotational variants~\cite{jeyaj_equiv}. In this work, we limit our focus to the well-known PLP and the PSP.

When it comes to modeling the vehicles on streets, the use of Poisson point processes (PPPs) is well established, where a street system is formed by PLP or PSP and independent 1D PPPs are placed on each street, forming Cox vehicular networks. We refer to the PLP-based and PSP-based vehicular point processes as the PLP-PPP and PSP-PPP, respectively. The selection of transmitters is governed by the channel access schemes used in vehicular networks. Ideally, the transmitting vehicles should be distributed such that no other transmitting vehicle is in its vicinity, mimicking a hard-core point process. However, the analysis of hard-core models is less tractable~\cite{koufos2}. In this work, we assume that the vehicles transmit with a certain probability in each slot following slotted ALOHA, {\em{i.e.,}} the transmitting vehicles on each street form independently thinned 1D PPPs. The PPP-based models essentially provide lower bounds on the performance of hard-core models~\cite{tong}. 

\subsection{Motivation}
Although the PPPs are tractable, coupling them with PLPs/PSPs turns them into Cox models that lead to unwieldy analytical expressions for the success probability~\cite{jeyaj_equiv, chetlur}. On the other hand, vehicle locations cannot be simply modeled as random points as in a 2D PPP neglecting the street geometry~\cite{qimei}. The random locations of points in a PPP defy the certainty of vehicles to be located on a line/stick. Hence there is a need for an intermediate model that is not as complicated as the Cox vehicular networks but not oversimplified as a 2D PPP, {\em{i.e.,}} a model that provides high tractability {\em and} good accuracy. Such a model should be effective not only to evaluate the success probability but also to evaluate more refined metrics such as the SIR meta distribution (MD), where SIR is the signal-to-interference ratio.  

The SIR MD naturally includes a reliability constraint on the individual transmitter-receiver links~\cite{md_main, md_letter1, md_letter2}. It answers questions like \textit{what fraction of the links support a target data rate of $10$ Kbps with a probability of at least $0.99$?}, whereas the success probability answers questions like \textit{what fraction of the links support a target data rate of $10$ Kbps?}, focusing on the average performance without reliability constraint. Note that the success probability is a special case of the SIR MD, in a precise sense defined later. 

\subsection{Related Work}
The success probability of the typical general vehicle receiving from a transmitter at a certain distance is analyzed for the case of the PLP-PPP in~\cite{chetlur, jeyaj_equiv}, and the PSP-PPP in~\cite{jeyaj_equiv}. The typical general vehicle reflects the average performance of all the vehicles. For the same setup in~\cite{chetlur}, a more comprehensive expression is derived in~\cite{jeyaj_equiv} that also comprises the success probability of the typical intersection vehicle, which corresponds to the average performance of the vehicles that are at intersections.  
Another line of work involving the PLP-PPP focuses on the typical general vehicle successfully receiving a message from its nearest neighbor, which can be either another vehicle or an infrastructure node such as a roadside unit or a cellular base station~\cite{morlot, dhillon, sial, choi2, chetlur_recent}. The transmitting and receiving vehicles on each street are modeled as independent 1D PPPs. The roadside units are placed on the streets following a 1D PPP. The base stations form an independent 2D PPP. 

The SIR MD, the main metric used in this paper, was first introduced in~\cite{md_main} and was evaluated for Poisson bipolar and cellular networks. Further, it was extended to carry out a fine-grained analysis of the base station cooperation, power control, and device-to-device (D2D) underlays in cellular networks. Apart from the SIR, the MD can also be defined for the data rate, energy harvested, etc. (see~\cite{deng} and references therein). The SIR MD for a vehicular network formed around an intersection is studied through simulations in~\cite{henk_meta}. The intersection is formed by two finite road segments with vehicles forming a PPP on each segment. It is shown that the MD is bimodal, {\em{i.e.,}} the individual link success probabilities are either low or high, not concentrated near their average. The MD of the SIR in linear motorways is analyzed in~\cite{koufos_meta} using a model where the inter-vehicle distances follow the shifted-exponential probability density function. 

Furthermore, using the SIR MD, we can find the maximum density of concurrently active links that satisfy a certain reliability constraint, referred to as the spatial outage capacity (SOC)~\cite{soc}. The SOC captures the trade-off between the density of active links and the fraction of the reliable links. Also, we can find insights on how to adapt the transmission parameters at a given vehicle density to avoid network congestion. When the channel load increases beyond a certain threshold, the number of packet collisions sharply increases, and the channel is said to be congested. The common methods to combat congestion include controlling the transmit i) rate, ii) power, and iii) data rate, and their combinations~\cite{liu}. In this work, we show how to handle congestion using the SIR MD by exploring the trade-off between the transmit rate and the fraction of reliable links.  

\subsection{Contributions}
We take the middle route between the complicated Cox vehicular network models and the oversimplified 2D PPP and introduce a model that provides a good trade-off between accuracy and tractability. The model, termed the transdimensional Poisson point process (TPPP), is the superposition of one or two 1D PPPs and a 2D PPP. By such superposition, we account for the geometry of the street(s) passing through the receiver, and, at the same time, we obviate the need to incorporate the geometry of the remaining streets. The main contributions are:
\begin{itemize}
\item[1.] We derive the SIR meta distribution for the PLP-PPP, and we show that it can be well approximated by that for the TPPP. In particular, the approximation is tight in the asymptotic regimes of data rate and reliability. We establish that the TPPP is also sufficient to capture the complex characteristics of the PSP-PPP.
\item[2.] We prove that shadowing further improves the accuracy of the TPPP. Precisely, the performance gap between the Cox vehicular networks and their TPPPs is the highest when there is no shadowing and vanishes as the shadowing variance increases. Notably, the maximum difference between the success probabilities in the PLP-PPP and TPPP in the case of no shadowing is about -14 dB.
\item[3.] Lastly, we introduce an SIR MD-based congestion control scheme that provides insights into adapting the transmit rate to guarantee a target fraction of reliable links. The advantage of the SIR MD is that it ensures each link is reliable with at least a probability of $x$ thus achieving fairness among the vehicles. In contrast, the success probability, which is the average reliability of the links, is not a suitable metric for network congestion since it cannot ensure the per-link performance. To the best of our knowledge, this is the first paper that uses the SIR MD as the target metric for adapting the transmit rate. 
\end{itemize}

\section{Vehicular Network Modeling and SIR Meta Distribution}
We will use the definitions and notations presented in this section throughout the paper unless otherwise stated. Let $o \triangleq (0,0)$ indicate the origin. Let $b(x,r)$ refer to a disk of radius $r$ centered at $x$ and $\vert \cdot \vert_{d}$ denote the Lebesgue measure in $d$ dimensions. 
\subsection{Modeling Streets and Vehicle Locations}
\label{sec:background}
A street system $\mathcal{S}$ is defined as the union of 1D subsets such as lines, line segments/sticks, etc., with no singletons or isolated points~\cite{jeyaj_equiv}. Here we present the formulations of PLP-based and PSP-based street systems.  

A line ({\em{i.e.,}} an infinitely long street) $L$ in $\mathbb{R}^{2}$ is denoted as
\begin{equation}
L({t},{\varphi_{t}}) = \lbrace{(x,y) \in \mathbb{R}^{2} : x\cos\varphi_{t} + y\sin\varphi_{t} = t \rbrace},
\label{eq:line}
\end{equation}
where $\vert t \vert$ is the distance of the perpendicular from the origin $o$ to the line, and $\varphi_{t}$ is the angle the perpendicular makes with the positive $x-$axis. Let $\mathcal{P}$ denote a 1D PPP of intensity $\mu$ and $\varphi_{t}$ be i.i.d. on $[0, \pi)$. The random countable collection of lines $\Xi_{\mathrm{L}} = \lbrace L(t, \varphi_{t}) : t \in \mathcal{P}, \varphi_{t} \in [0, \pi) \rbrace$ forms the PLP. The PLP-based street system is the union of lines, {\em{i.e.,}} $\mathcal{S} = \bigcup_{L \in \Xi_{\mathrm{L}}} L$.

A stick $S(y, \varphi_{y}, h_{y})$ of orientation $\varphi_{y}$ and half-length $h_{y}$, centered at the midpoint $y$ is formally defined as
\begin{align}
S(y, \varphi_{y}, h_{y}) = y + {\mathrm{rot}}_{\varphi}([-h_{y},h_{y}]), 
\label{eq:segment}
\end{align}
where $y$ follows a 2D PPP of intensity $\mu$, $\varphi_{y}$ is i.i.d. on $[0, \pi)$, and $h_{y}$ is i.i.d. with some density function $f_{H}(h)$. $\mathrm{rot}_{\varphi}$ denotes the rotation of stick by $\varphi_{y}$ around $o$. Let $\Xi_{\mathrm{S}} = \lbrace S(y, \varphi_{y}, h_{y}) \rbrace $ denote the collection of sticks that forms the PSP. Then $\mathcal{S} = \bigcup_{S \in \Xi_{\mathrm{S}}} S$ denotes the PSP-based street system. We alternately write $y$ in polar coordinates as $(\gamma, \phi_{y})$. Now, we are ready to define the vehicular point processes with respect to $\mathcal{S}$.
\begin{figure*}
\centering
\subfloat[]{%
\includegraphics[scale = 0.7]{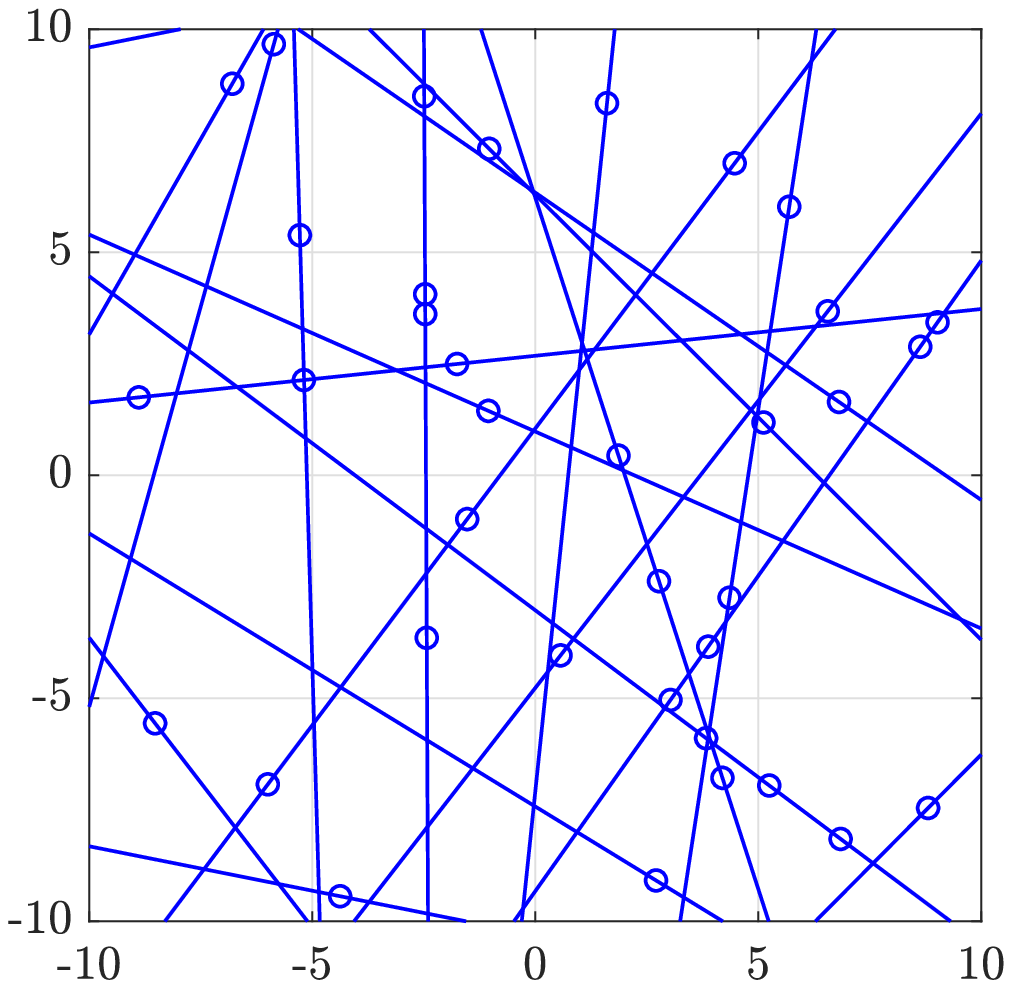}
\label{fig:plp_ppp}
 }\hspace{10mm}
\subfloat[]{%
\includegraphics[scale = 0.7] {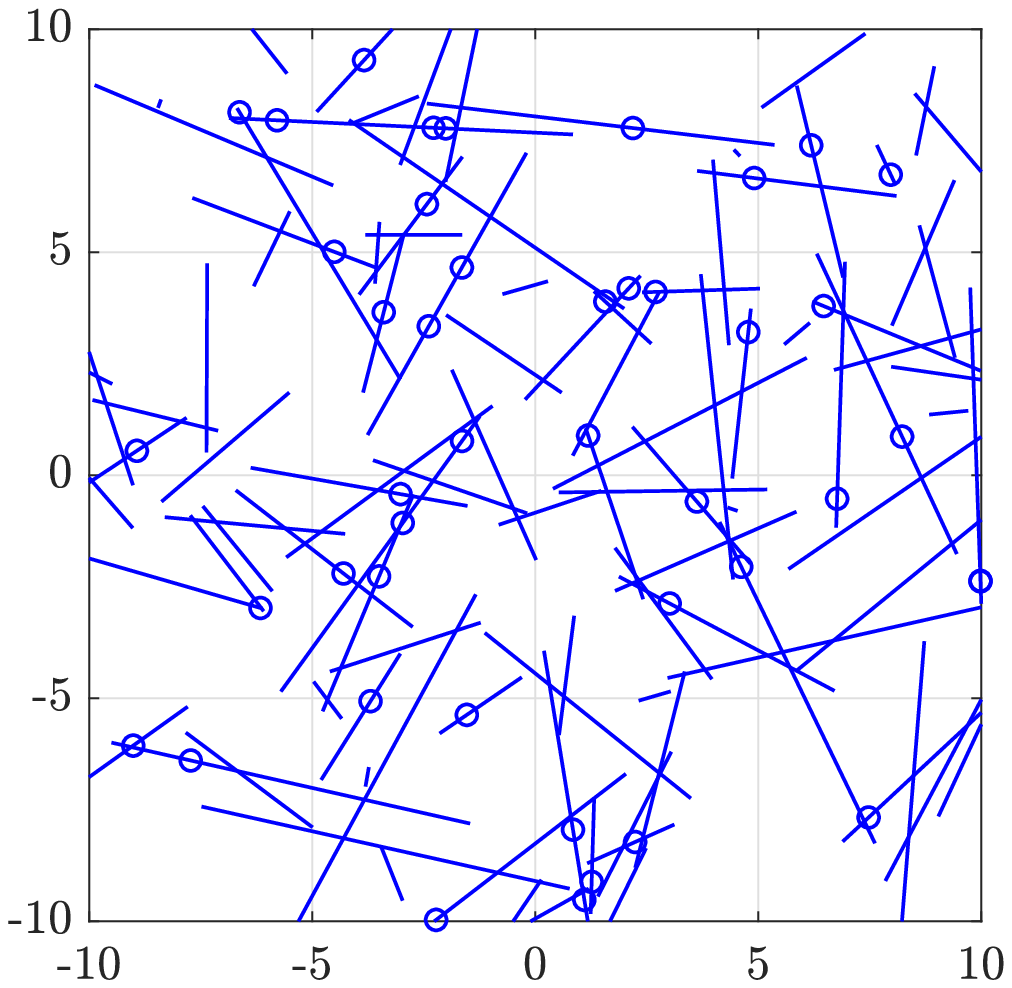}
\label{fig:psp_ppp}
 }
\caption{\label{fig:models} Snapshots of the (a) PLP-PPP and (b) PSP-PPP. Lines or sticks denote the streets, and `$\circ$' denote the vehicles. $\mu = 0.1$, $\lambda = 0.1$, and $f_{H}(h) = 2 c h \exp(-ch^{2})$ with $c = 0.1$. }
\end{figure*}

A vehicular point process $\mathcal{V} \subset \mathbb{R}^2$ is a Cox process with random intensity measure $\Upsilon(\mathcal{B})=\lambda |\mathcal{S}\cap \mathcal{B}|_1$. This implies that the vehicles  on each street form independent 1D PPPs of intensity $\lambda$. For a stationary $\mathcal{S}$, we have 
\setcounter{equation}{2}
 \begin{equation}
\mathbb{E}[\vert \mathcal{S} \cap \mathcal{B} \vert_1] = \tau \vert \mathcal{B} \vert_2 \hspace{3mm} \text{for Borel sets} \hspace{1mm} \mathcal{B} \subset \mathbb{R}^2, 
\label{eq:street}
\end{equation}
where $\tau$ is the mean total street length per unit area. Then, the 2D intensity measure of $\mathcal{V}$ is $\mathbb{E}[\Upsilon(\mathcal{B})]=\lambda \mathbb{E}[|\mathcal{S}\cap \mathcal{B}|_1] = \lambda \tau\vert\mathcal{B}\vert_2$. The vehicular point processes based on the PLP and PSP are referred to as the PLP-PPP and PSP-PPP, respectively. Fig.~\ref{fig:models} depicts their sample realizations.

The PLP and PSP inherently form intersections. A vehicle at $z \in \mathbb{R}^{2}$ is of order $m$ if 
\begin{equation}
\vert \mathcal{S}\cap b(z,r)\vert_1 \sim mr,\quad r \to 0 . 
\end{equation}
This implies that the vehicles at endpoints are of order $1$, those at intersections are of order 4, and those at all other locations are of order 2.

\subsection{Properties}
Below, we list a few properties of the PLP-PPP and PSP-PPP, and PPPs that we will need in the rest of the paper. Let $\Phi_d$ denote a homogeneous $d$-dimensional PPP of intensity $\lambda_{d}$. Let $c_{d}$ denote the volume of the unit $d-$dimensional ball, {\em{i.e.,}} $c_{1} = 2$ and $c_{2} = \pi$. 

\begin{lemma} [\hspace{-1.2mm}  \cite{jeyaj_equiv}]
The mean total street length per unit area in the PLP-PPP is $\tau = \mu$. In the PSP-PPP, it is $\tau = 2 \mu \mathbb{E}[H]$. \label{lemma_lsp_2d}
\end{lemma}

\begin{lemma} [\hspace{-1.5mm}  \cite{haenggi, jeyaj_equiv}]
The nearest-neighbor distance distributions for the PPP $F_{R}^{\Phi_d}(r)$, PLP-PPP $F^{\mathrm{PLP-PPP}}_{R}(r)$, and PSP-PPP $F^{\mathrm{PSP-PPP}}_{R}(r)$ are given by
\begin{align}
F_{R}^{\Phi_d}(r) & = 1-\exp(-c_{d} \lambda_{d} r^{d}),
\label{eq:nnf_nd_ppp} \\
F^{\mathrm{PLP-PPP}}_{R}(r) & = 1-\exp (-\lambda  m r - 2 \mu \smallint_{0}^{r} (1-\exp(-2 \lambda \sqrt{r^2-u^2}\hspace{0.3mm}) \mathrm{d}u ), \label{eq:nnf_plp} \\
F^{\mathrm{PSP-PPP}}_{R}(r) & = 1 - \bigg[\int \limits_{0}^{\infty} \frac{1}{h}\int \limits_{0}^{h} \exp(-\lambda \ell(r, \gamma, 0, 0, h) )   \tilde{f}_{H}(h)  \hspace{0.1mm}  \mathrm{d}\gamma  \hspace{0.1mm}  \mathrm{d}h \bigg]^{m/2} \nonumber \\
& \hspace{8mm} \times \exp\bigg(-\frac{\mu}{\pi}\int \limits_{0}^{\infty} \int \limits_{0}^{\pi} \int \limits_{0}^{2\pi} \int \limits_{0}^{r+h}   \exp(-\lambda \ell(r, \gamma, \phi, \varphi, h)) \gamma  {f}_{H}(h) \hspace{0.1mm}\mathrm{d}\gamma \hspace{0.1mm}  \mathrm{d}\phi \hspace{0.1mm} \mathrm{d}\varphi \hspace{0.1mm}  \mathrm{d}h \bigg),
\label{eq:nnf_psp}
\end{align}
respectively. The term $\ell (r, \gamma, \phi,\varphi, h)$ in \eqref{eq:nnf_psp} equals $\ell_1  (\gamma, \phi,\varphi, h) 
\mathds{1}_{\gamma \leq r} +  \ell_2(r, \gamma, \phi,\varphi, h)  \mathds{1}_{\gamma > r} $, where $\ell_{k}(r, \gamma, \phi,\varphi, h) = \min(u_{k}(r, \gamma, \phi,\varphi), h) - (-1)^{k} \min(u_{k}(r, \gamma, \phi,\varphi), h)$ with $u_{k}(r, \gamma, \phi,\varphi)  = \vert -\gamma \cos(\phi-\varphi) - (-1)^k \sqrt{r^2-\gamma^{2}\sin^{2}(\phi-\varphi)} \vert$ for $k \in \lbrace 1, 2 \rbrace$, and $m \in \lbrace{2,4 \rbrace}$.
\label{nnf_plp_psp}
\end{lemma}

\subsection{Communication Model} Each vehicle broadcasts with probability $p$ in each time slot following slotted ALOHA. Thus the intensity of active transmitters on each street is $\lambda p$. We focus on the probability that a vehicle successfully receives the message from a transmitter at distance $D$. If a vehicle can successfully receive the message over a distance $D$, then all the other receivers within distance $D$ are likely to receive the message, too. This model can be extended to analyze any form of vehicle-to-everything (V2X) communication since the transmitter is only specified by its distance $D$ to the typical vehicle at the origin. The transmitter can be a vehicle, a roadside unit, a pedestrian, or any other infrastructure node, {\em{i.e.,}} the transmitter is the `X' in V2X.

The performance metrics are calculated with respect to the typical vehicle. We condition a vehicle (receiver) to be at the origin $o$. On averaging over the point process, this vehicle becomes the typical vehicle. For a stationary street system $\mathcal{S}$, we can condition the vehicle to be at an arbitrary location. As vehicles are located on the streets, having a vehicle at $o$ implies that at least one street passes through $o$. The typical vehicle of order $m$ has $ m/2 $ streets passing through $o$. The typical vehicle of order $2$ is referred to as the typical general vehicle, and that of order $4$ is the typical intersection vehicle. The term `typical vehicle' without qualification refers to both kinds of vehicles.

The typical vehicle's transmitter is assumed to be active and located at a distance $D$ from the origin. The SIR at the typical vehicle at the origin is given by
\begin{equation}
{\sf SIR} = \frac{g D^{-\alpha}}{\sum_{z \in \mathcal{V}} g_{z} \Vert z \Vert^{-\alpha} B_{z}}.
\label{eq:sir_eqn}
\end{equation}
The denominator in~\eqref{eq:sir_eqn}, $I = \sum_{z \in \mathcal{V}} g_{z}\Vert z \Vert^{-\alpha}$, is the total interference power at the origin. The channel power gain $g$ is exponentially distributed with mean 1 (Rayleigh fading), $\alpha $ is the path-loss exponent, and the $B_{z}$'s are i.i.d. Bernoulli random variables with mean $p$.

We partition the vehicular point process $\mathcal{V}$ as $\mathcal{V} = \mathcal{V}_{o}^{m} \cup \mathcal{V}_!$, where $\mathcal{V}_{o}^{m}$ and $\mathcal{V}_! $ are the point processes of the vehicles on the streets that pass through the typical vehicle of order $m$ and on the remaining streets, respectively. Accordingly, we can write the total interference $I$ as $I = I_{o}^{m} + I_{\mathrm{!}}$, where $I_{o}^{m}$ and $I_{\mathrm{!}}$ denote the interference from the transmitting vehicles on the typical vehicle's streets and remaining streets, respectively. Further, let $\delta = 2/\alpha$. Table~\ref{tab:symbols} lists the frequently used acronyms and notations in the paper.

\begin{table}
\centering
\caption{Acronyms and Notations}
\begin{tabular}{|r|l|}
  \hline
  PPP & Poisson point process \\
  PLP & Poisson line process \\
  PSP & Poisson stick process \\
  TPPP & Transdimensional Poisson point process \\
  SIR & Signal-to-interference ratio \\
  MD & Meta distribution \\ 
  $\mathcal{S}$ & Street system \\  
  $\mu$ & Street intensity \\
  $\lambda$ & Vehicle intensity on each street \\
  $H$ & Half-length of the street in the PSP \\ 
  $D$ & Distance between the typical vehicle and its transmitter \\
  $m$ & Order of the typical vehicle \\  
  $p$ & Transmit probability \\
  $\theta$ & SIR threshold \\ 
  $P_{m}$ & Conditional success probability \\
  $p_{m}$ & Success probability \\
  $M_{b,m}$ & $b$th moment of $P_{m}$ \\
  $\bar{F}_{P_{m}}$ & SIR MD \\
  $x$ & SIR MD reliability threshold \\
  $\alpha$ & Path-loss exponent \\ 
  $\mathcal{V}$ & Vehicular point process \\
  $\mathcal{V}_{o}^{m}$ & Vehicular point process on the typical vehicle's streets \\
  $I$ & Total interference \\
  $I_{o}^{m}$ & Interference with respect to $\mathcal{V}_{o}^{m}$ \\ 
  $\Phi_{d}$ & $d-$dimensional PPP \\
  $\lambda_{d}$ & Intensity of $\Phi_{d}$ \\
  \hline
\end{tabular}
\label{tab:symbols}
\end{table}

\subsection{SIR Meta Distribution}
The conditional success probability of the typical vehicle of order $m$ is 
\begin{equation}
P_{m}(\theta) = \mathbb{P}({\sf{SIR}} > \theta \mid \mathcal{V}),
\label{eq:link_ps}
\end{equation}
where $\theta$ is the SIR threshold that parametrizes the data rate. Conditioning on $\mathcal{V}$ in~\eqref{eq:link_ps} implies that we average only over the fading and slotted ALOHA.
The meta distribution of the SIR is given by~\cite{md_main, md_letter1}
\begin{equation}
\bar{F}_{P_{m}}(\theta, x) = \mathbb{P}(P_{m}(\theta) > x), \hspace{3mm} x \in [0,1], 
\label{eq:md}
\end{equation}
where $x$ is the reliability threshold.  
By the Gil-Pelaez theorem, the SIR meta distribution can be expressed as
\begin{align}
\bar{F}_{P_{m}}(\theta, x) = \frac{1}{2} + \frac{1}{\pi} \int_{0}^{\infty} \frac{\Im(e^{-jt \log{x}}M_{jt, m})}{t} \mathrm{d}t,
\label{eq:exact_meta}
\end{align}
where
\begin{equation}
M_{b, m}(\theta) = \mathbb{E}[P_{m}(\theta)^{b}], \hspace{2mm} b \in \mathbb{C}.
\label{eq:mb_formula}
\end{equation}
The average of the conditional success probabilities is the success probability $p_{m}$, {\em{i.e.,}}
\begin{align}
p_{m} = \mathbb{E}[P_{m}(\theta)] = \mathbb{P}({\sf{SIR}} > \theta). \label{eq:psuc} 
\end{align}
Since $p_{m} = M_{1,m}(\theta)$, the terms `first moment' and `success probability' can and will be used interchangeably.

\section{The Transdimensional Poisson Point Process}
We propose a transdimensional model that includes the vehicles on the street(s) passing through the typical vehicle of order $m$ at the origin and models the remaining vehicles on the plane as a 2D PPP neglecting the geometry of the other streets. The formal definition follows.  

\begin{definition}
\label{def:TPPP}
Let $\Psi_k = \lbrace(t_1, 0), (t_2, 0), \dotsc\rbrace$ where $\lbrace t_{i}\rbrace$, $i \in \mathbb{N}$, is a 1D PPP of intensity $\lambda$ on $\mathcal{R}_{k} \subseteq \mathbb{R}$, $1 \leq k \leq m/2 $, and $m \in \lbrace 2, 4 \rbrace$. Let $\Psi_{o}^{m} = \bigcup \limits_{k = 1}^{ m/2} \Psi_{k}$ denote the point process on $ m/2$  streets that pass through the typical vehicle of order $m$. The transdimensional Poisson point process (TPPP) with respect to the typical vehicle of order m is the superposition of $\Psi_o^{m}$ and a 2D PPP $\Phi_2$ of intensity $\lambda_2$, i.e., $\mathcal{T} \triangleq \Psi_o^{m} \cup \Phi_2$. 
\end{definition}

\begin{figure*}
\subfloat[]{\includegraphics[scale=.47]{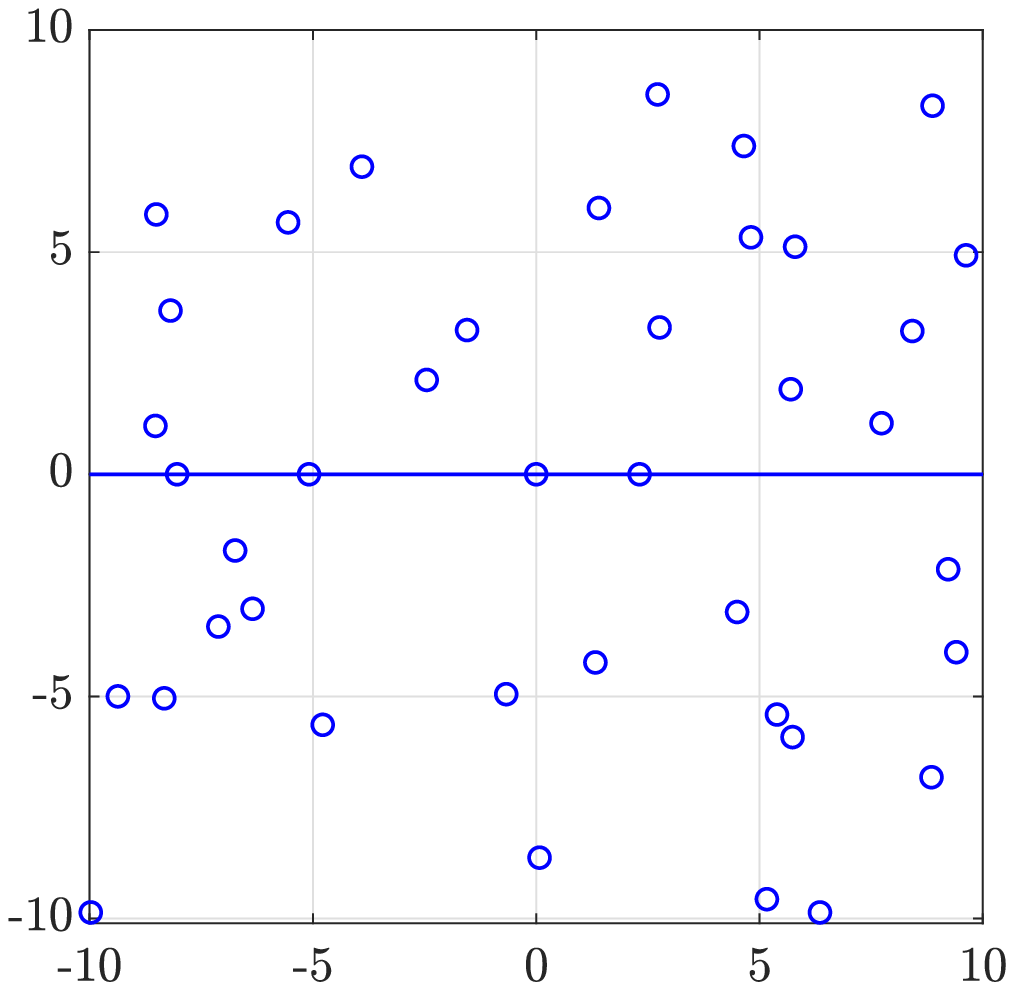}{\label{fig:tppp_models}}}\hfill
\subfloat[]{\includegraphics[scale=.47]{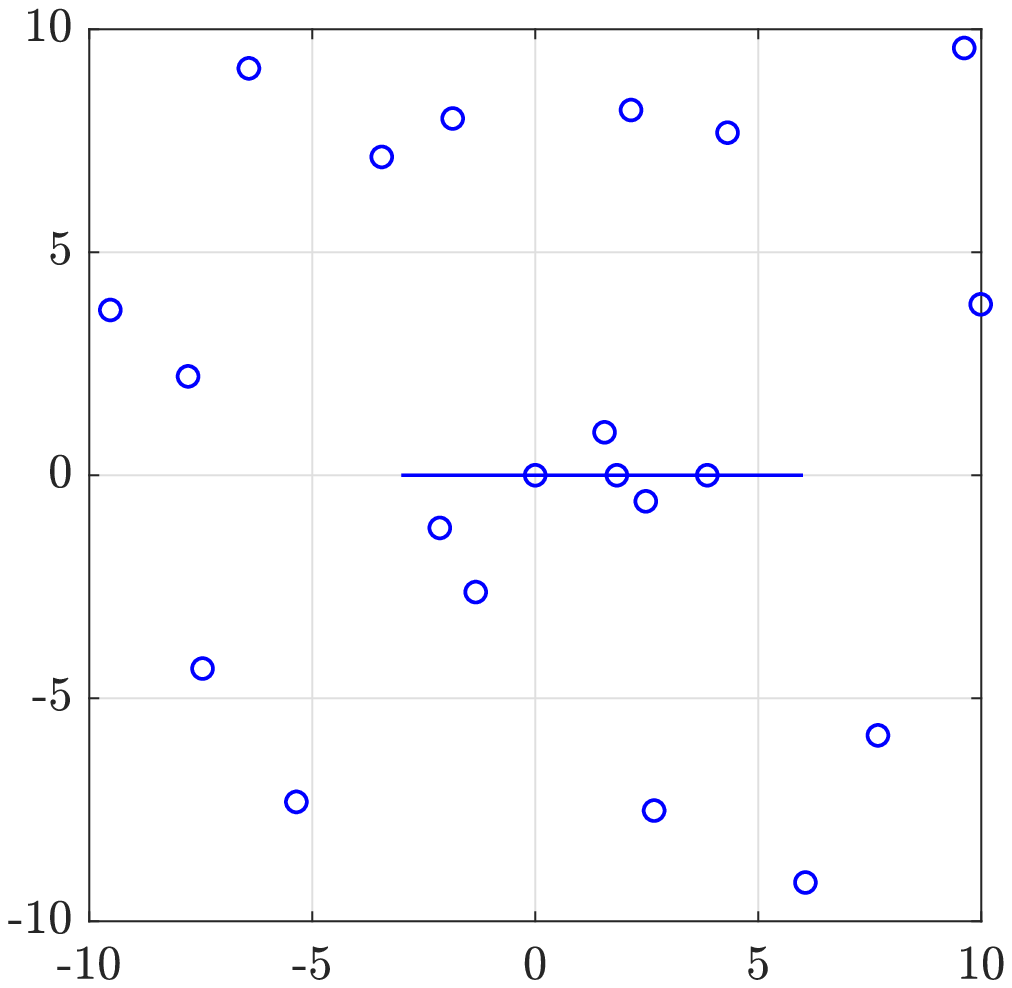}{\label{fig:ps_stppp_gen}}}\hfill
\subfloat[]{\includegraphics[scale=.47]{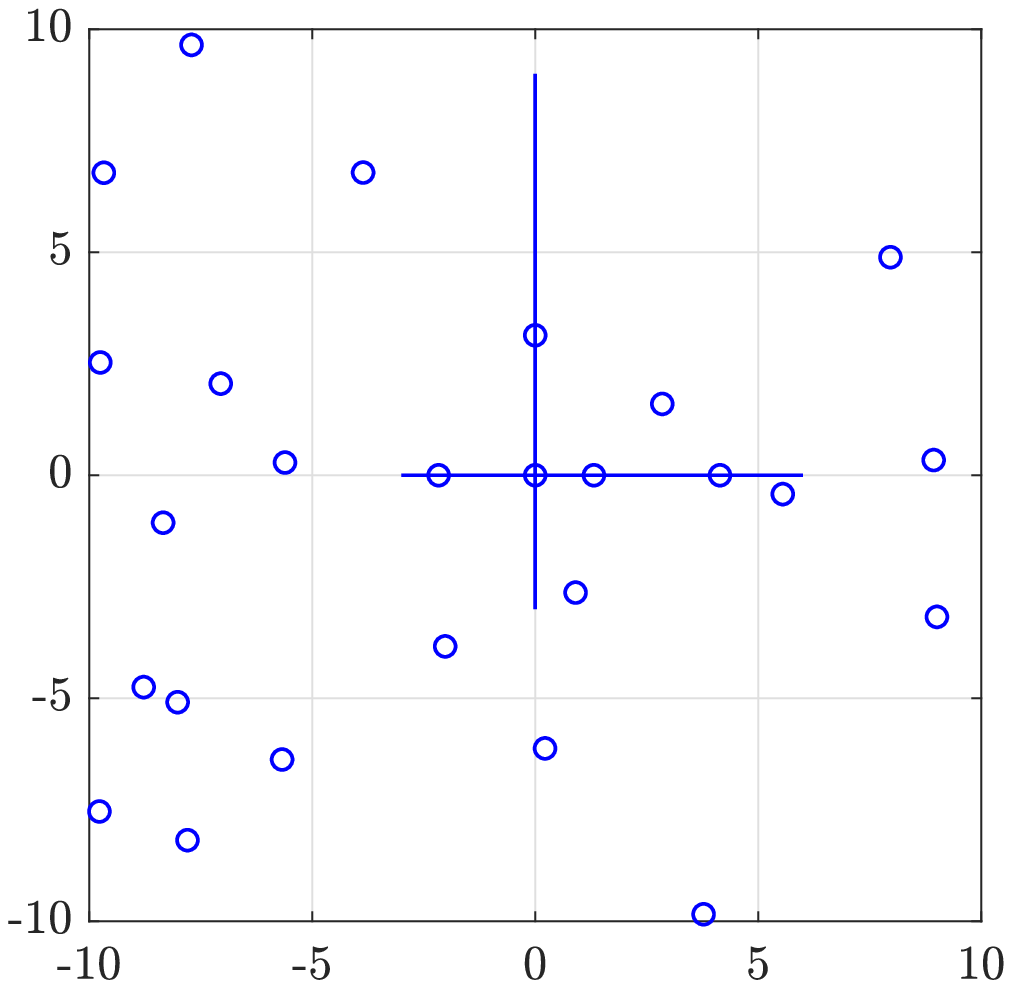}{\label{fig:ps_stppp_inter}}}
\caption{\label{fig:stppp} {Snapshot of the TPPP with respect to the typical vehicle in the PLP-PPP is shown in (a), where $m\lambda/2 = \lambda_2 = 0.1$. Snapshots of the TPPP with respect to the typical general and intersection vehicles in the PSP-PPP are shown in (b) and (c), respectively, where $\lambda = 0.3$, $\mu = 0.01$, $f_{H}(h) = 2ch\exp(-ch^{2})$ with $c = 0.01$, and $\lambda_2 = 2 \lambda \mu \mathbb{E}[H]$. Line/stick denotes a street and `o' denotes a vehicle. }}
\end{figure*}

$\Psi_o^{m} $ is equivalent in distribution to $\mathcal{V}_{o}^{m}$ as the 1D PPPs on $m/2$ streets are independent, and their orientations are i.i.d. $\mathcal{R}_{k}$ is an infinite or a finite  interval that contains the origin. $\mathcal{R}_{k} = \mathbb{R}$ if the street is a line, and $\mathcal{R}_{k} \subset \mathbb{R}$ if the street is a stick. By the superposition property of the PPP, $\Psi_o^{m}$---the union of 1D PPPs of intensity $\lambda$ on $m/2$ lines passing through the typical vehicle of order $m$---is equivalent to a 1D PPP of intensity $m\lambda/2$. By Lemma~\ref{lemma_lsp_2d}, $\lambda_2 = \lambda  \mu$. Fig.~\ref{fig:tppp_models} shows a realization of the TPPP corresponding to the PLP-PPP.

The superposition property does not extend to the PSP-PPP even if the sticks are of the same finite length. The reason is that the origin can be located at a distance $w \in (0,h_{y})$ from the midpoint $y$ of the stick. The lengths of the sticks on both sides of the origin need not be the same. In other words, the interference measured at the origin with respect to the two streets may differ. The streets passing through the origin form a Cox process as they are stochastic with respect to the length of the stick as well as their starting or ending points. Figs.~\ref{fig:ps_stppp_gen} and \ref{fig:ps_stppp_inter} show the snapshots of the TPPP with respect to the typical general and intersection vehicles in the PSP-PPP, where the half-length of a stick follows the Rayleigh distribution. For the sake of visualization, we show the streets to be orthogonal in Fig.~\ref{fig:ps_stppp_inter} rather than being on top of each other as given in Definition~\ref{def:TPPP}. We have $\lambda_2 = 2 \lambda \mu \mathbb{E}[H]$ by Lemma~\ref{lemma_lsp_2d}. 

The TPPP $\mathcal{T}$ is non-stationary as the neighborhood seen by a point in $\Psi_{o}^{m}$ is different from that in $\Phi_2$ as at least one street passes through the typical vehicle. 

Next, we analyze the SIR meta distributions for the PLP-PPP, PSP-PPP, and their respective TPPPs, and discuss the logic behind using the TPPP for vehicular network analysis. To this end, we derive the moments~\eqref{eq:mb_formula} required to calculate the SIR meta distribution~\eqref{eq:exact_meta} for the PLP-PPP and the corresponding TPPP; then, we perform a comparative analysis of the moments of different orders in the PLP-PPP and TPPP followed by that of their respective SIR meta distributions.

\section{The Transdimensional Approach to the PLP-PPP}
\label{sec:ltppp}
\subsection{Derivation of Moments}
\begin{theorem}
\label{md_plp_result}
The $b$-th moment of the conditional success probability $P^{\mathrm{PLP-PPP}}_{m}(\theta)$ of the typical vehicle of order $m \in \lbrace 2, 4 \rbrace$ in the PLP-PPP is given by
\begin{align}
M^{\mathrm{PLP-PPP}}_{b,m}  =   \exp (&-m \lambda D\theta^{\delta/2}  \Gamma(1+\delta/2) \Gamma(1-\delta/2)   \mathfrak{D}_{b}(p, \delta/2) -2  \mu \smallint _{0}^{\infty} ( 1-  G_{b}(t) ) \mathrm{d}t ), \label{eq:mb_plp}
\end{align}
where $\mathfrak{D}_{b}(p, \delta/2) = pb \hspace{0.5mm} _{2}F_{1} (1-b, 1-\delta/2; 2; p)$,  $\delta = 2/\alpha$,  $G_{b}(t) = \exp \bigg(- \lambda  \delta \displaystyle \int _{t^{2/\delta    }}^{\infty} \bigg[ 1- \bigg(1- \frac{p s}{v + s } \bigg)^{b} \bigg]\frac{v^{\delta -1}}{\sqrt{v^{\delta}-t^{2}}} \mathrm{d}v \bigg)$, and $s = \theta D^{\alpha}$.
\end{theorem}
\begin{IEEEproof}
See Appendix~\ref{appendix:md_plp_ppp}.
\end{IEEEproof}

\begin{corollary}
\label{lemma_plp}
The first moment $M^{\mathrm{PLP-PPP}}_{1,m}$ or, equivalently, the success probability $p^{\mathrm{PLP-PPP}}_{m}$ is 
\begin{align}
M^{\mathrm{PLP-PPP}}_{1,m} = \exp(&- m \lambda p D  \theta^{\delta/2} \Gamma(1+\delta/2)\Gamma(1-\delta/2) 
 - 2 \mu \smallint_{0}^{\infty} (1- \mathcal{L}_{I_t}(\theta D^{\alpha})) \hspace{0.3mm} \mathrm{d}t), \label{eq:ps_plp_ppp}
 \end{align}
 where $\mathcal{L}_{I_t}(s) = \exp \big(-  \lambda p  s^{\delta/2} \int_{u_t}^{\infty} \frac{1}{\left( 1 + u^{{1/\delta}}\right) \sqrt{u - u_t}} \mathrm{d}u\big)$ with $u_t = t^2s^{-\delta}$.
\end{corollary}
\begin{IEEEproof}
It directly follows from~\eqref{eq:mb_plp} by noting that $\mathfrak{D}_{1}(p, \delta/2) = p$ and the change of variable $u = v^{\delta}$ in $G_{b}(t)$ with $b = 1$. 
\end{IEEEproof}

See~\cite{jeyaj_equiv} and~\cite{chetlur} for alternative proofs.
\begin{corollary}
\label{cor_var0}
For a given transmitter density $\lambda p = C$, as $p \to 0$, we have \[ \lim_{\substack{p \to 0 \\ \lambda p = C}} P^{\mathrm{PLP-PPP}}_{m} = M^{\mathrm{PLP-PPP}}_{1, m}\]
in mean square (and probability and distribution).
\end{corollary}
\begin{IEEEproof}
See Appendix~\ref{appendix:var0}.
\end{IEEEproof}
The conditional success probabilities converge to their average only when $p \to 0$. Generally, the average gives very little information on the individual links. 

\begin{theorem}
The $b$-th moment of $P^{\mathrm{TPPP}}_{m}(\theta)$ of the typical vehicle in the TPPP is given by 
\begin{align}
M_{b,m}^{\mathrm{TPPP}}   =   \exp(& -  m \lambda  D\theta^{\delta/2}  \Gamma(1+\delta/2) \Gamma(1-\delta/2) \mathfrak{D}_{b}(p, \delta/2) \nonumber \\
&  -  \lambda  \mu \pi D^{2} \theta^{\delta}  \Gamma(1+\delta) \Gamma(1-\delta) \mathfrak{D}_{b}(p, \delta)), \label{eq:mb_tppp}
\end{align}
where $\delta = 2/\alpha$ and $\mathfrak{D}_{b}(p, q) = pb \hspace{0.7mm} _{2}F_{1} (1-b, 1-q; 2; p)$.
\end{theorem}
\begin{IEEEproof}
In a $d-$dimensional PPP $\Phi_{d}$ of intensity $\lambda_d$, the $b$-th moment of the conditional success probability for a link distance $D$ is~\cite[Eqn. (4)]{md_main} 
\begin{equation}
M_{b}^{\Phi_{d}}  = \exp(-  \lambda_{d} c_{d} D^{d}\theta^{\delta'}  \Gamma(1+\delta') \Gamma(1-\delta') \mathfrak{D}_{b}(p, \delta')),
\label{eq:mb_nd_ppp}
\end{equation}
where $\delta' = d/\alpha$. By the independence of the point processes $\Psi_{o}^{m}$ and $\Phi_{2} \subset \mathcal{T}$, $M^{\mathrm{TPPP}}_{b,m}$ is the product of the moments of the conditional success probabilities in  $\Psi_{o}^{m}$ and $\Phi_{2}$ given by~\eqref{eq:mb_nd_ppp} with $\lambda_1 = (m/2)\lambda$ and $\lambda_2 = \lambda \mu$.
\end{IEEEproof}

The success probability $p_{m}^{\mathrm{TPPP}}$ of the typical vehicle is the first moment $M_{1,m}^{\mathrm{TPPP}}$. It is obtained by setting $b=1$ in~\eqref{eq:mb_tppp}, {\em{i.e.,}}
\begin{align}
M_{1,m}^{\mathrm{TPPP}} =  \exp(& -  m \lambda p  D\theta^{\delta/2}  \Gamma(1+\delta/2) \Gamma(1-\delta/2)  -  \lambda  p \mu \pi D^{2} \theta^{\delta}  \Gamma(1+\delta) \Gamma(1-\delta) ). \label{eq:m1_tppp}
\end{align} 

The second term in the exponential in~\eqref{eq:mb_plp} has two nested integrals over infinite ranges, while that in~\eqref{eq:mb_tppp} has none. The numerical evaluation of~\eqref{eq:mb_plp}, in particular when used in~\eqref{eq:exact_meta}, is tedious since it involves three layers of complex-valued integrals. Note that it takes about 100,000 times longer to numerically evaluate the first moment for the PLP-PPP~\eqref{eq:ps_plp_ppp} than that for the TPPP~\eqref{eq:m1_tppp}. To demonstrate that the TPPP is an accurate yet simple model, we compare the SIR MDs of the PLP-PPP and the corresponding TPPP, starting with the first moment.

\subsection{First-Order Moment Analysis for the PLP-PPP}
The moments of the conditional success probability for the PLP-PPP~\eqref{eq:mb_plp} do not have closed-form expressions. To gain insights into $M^{\mathrm{PLP-PPP}}_{1,m}$ or $p^{\mathrm{PLP-PPP}}_{m}$, we begin with the asymptotic analysis with respect to $\theta$. 

\begin{theorem} [{{{\hspace{-0.1mm}\cite{jeyaj_tppp}, Th. 4}}}] 
\label{th_zero}
The success probability of the typical vehicle tends to that in a  1D PPP as $\theta \to 0$, i.e.,
\begin{equation}
1-p_{m}^{\mathrm{PLP-PPP}}(\theta) \sim m\lambda p D {\theta^{\delta/2}} \Gamma(1+\delta/2) \Gamma (1-\delta/2), \hspace{2mm} \theta \to 0.
\end{equation}
\end{theorem}
Intuitively, as $\theta \to 0$, for SIR $> \theta$ to hold, it suffices not to have any interferers within a small disk around the typical vehicle. With high probability, the small disk intersects only the street(s) passing through the typical vehicle. Consequently, as $\theta \to 0$, the success probability of the typical vehicle in the PLP-PPP converges to that in the network formed only by the typical vehicle's streets.
\begin{theorem}[{{{\hspace{-0.1mm}\cite{jeyaj_tppp}, Th. 5}}}] 
\label{th_infty}
The success probability of the typical vehicle tends to that in a  2D PPP as $\theta \to \infty$, i.e.,
\begin{equation}
p_{m}^{\mathrm{PLP-PPP}}(\theta) \sim \exp(\hspace{-0.5mm}-  \pi \lambda p \mu D^{2} \theta^{\delta}\Gamma(1+\delta)\Gamma(1-\delta)), \hspace{2mm} \theta \to \infty. \label{eq:asym_infty}
\end{equation}
\end{theorem}
Here the intuition is that as $\theta \to \infty$, for SIR $> \theta$, a large disk around the typical vehicle must be devoid of interferers. The fact that $p_{m}^{\mathrm{PLP-PPP}}$ tends to the success probability of the typical vehicle in a 2D PPP as $\theta \to \infty$ signifies that the geometry of the vehicle locations outside the large disk does not matter. This is further corroborated by the nature of the pair correlation function of the PLP-PPP given by~\cite[Ch. 8]{stoyan}
\begin{equation}
g^{\mathrm{PLP-PPP}}(r)= 1 + \frac{1}{\mu r }.
\label{eq:pcf_plp}
\end{equation}
It tends to $1$ as $r \to \infty$ implying that the locations of the vehicles separated by larger distances are independent as in a PPP~\cite{haenggi}.

\begin{figure}
\centering
\subfloat[]{\includegraphics[scale=.6]{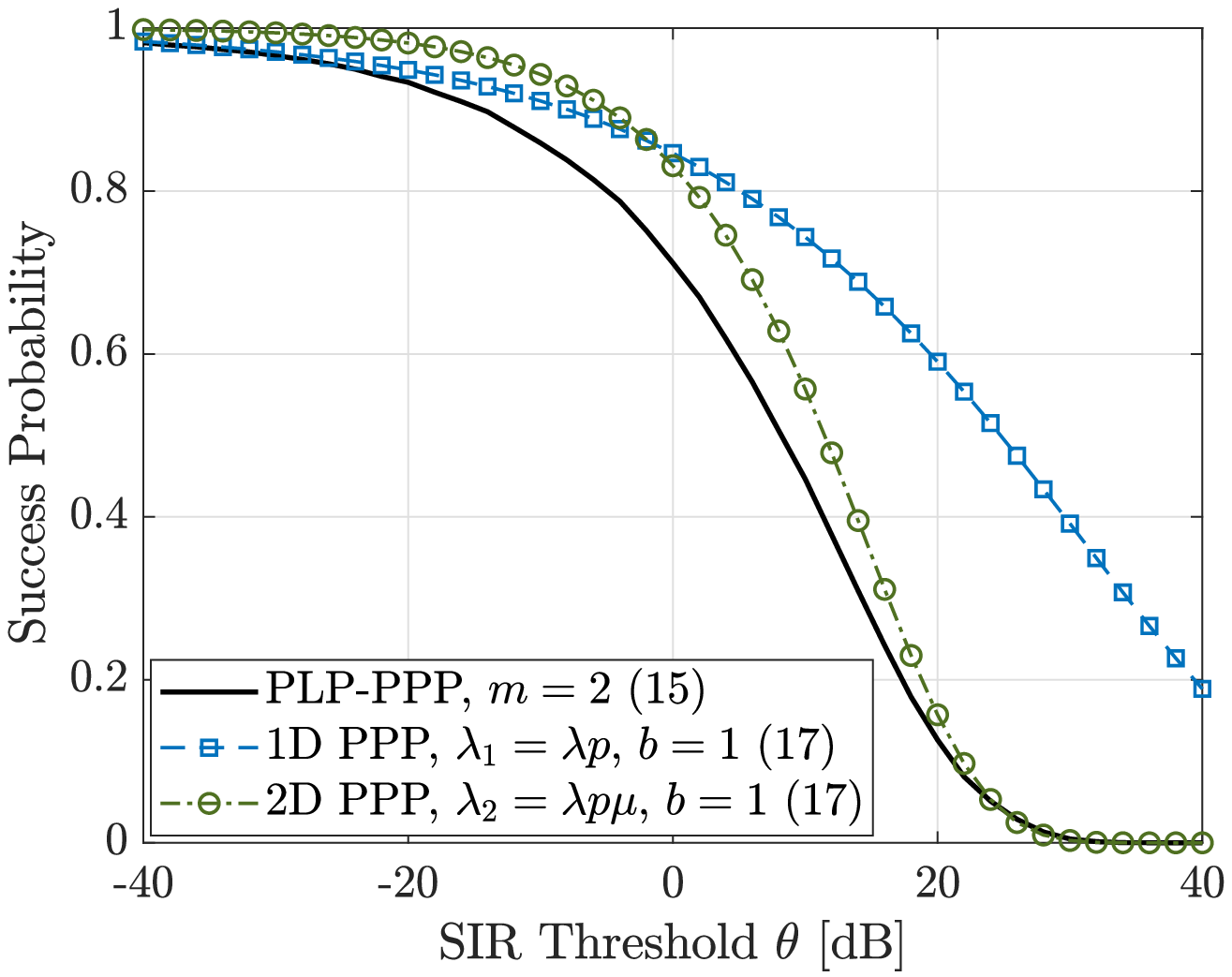}{\label{fig:ps_plp_ppp_gen}}}
\hfill
\subfloat[]{\includegraphics[scale=.6]{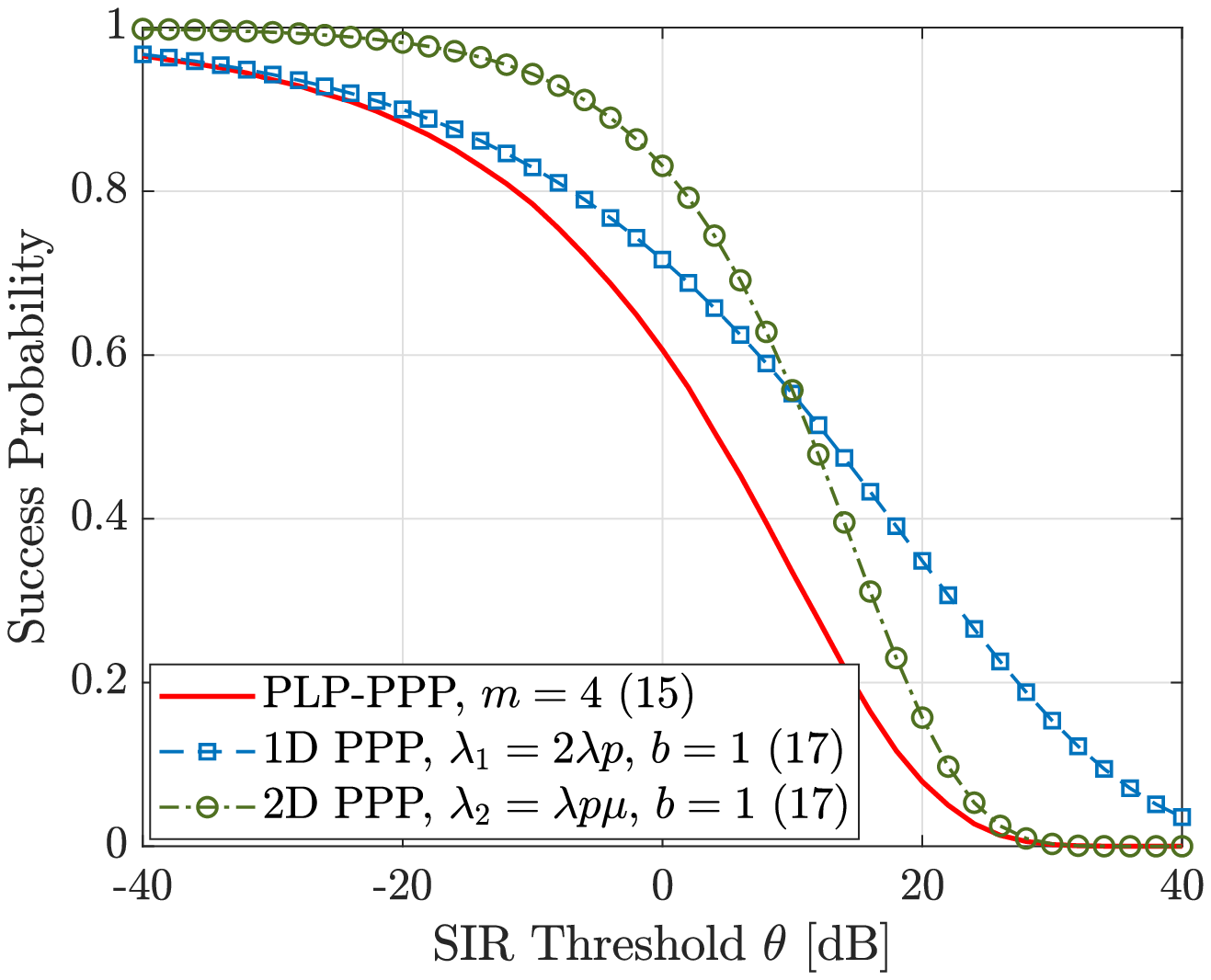}{\label{fig:ps_plp_ppp_inter}}}
\caption{\label{fig:ps_cox_ppp}{Comparison of success probabilities of the (a) typical general vehicle and (b) typical intersection vehicle in the PLP-PPP to that of the typical vehicle in 1D and 2D PPPs. $\mu = 2$, $\lambda = 1$, $p = 0.3$, $D = 0.25$, and $\alpha = 4$. The equation numbers are given in the parentheses in the legends.}}
\end{figure}
Fig.~\ref{fig:ps_cox_ppp} compares the success probability in the PLP-PPP to that in the 1D and 2D PPPs. We observe from Fig.~\ref{fig:ps_cox_ppp} that the success probability of the typical vehicle is upper bounded by the minimum of the success probabilities of the typical vehicle in 1D and 2D PPPs. This implies that either a 1D PPP or a 2D PPP alone is insufficient to characterize the vehicular network. We also see that the success probabilities of the typical vehicle in the PLP-PPP tend to that in the 1D and 2D PPPs in the asymptotic regimes as established in Theorems~\ref{th_zero} and \ref{th_infty}. This hints at the possibility of using a simpler, purely PPP-based model that has the properties of both the 1D and 2D PPPs for vehicular network analysis. In the next subsection, we show that the TPPP model indeed results in a highly accurate approximation of the PLP-PPP. 

\subsection{Comparison of First-Order Moments}
We remark that the TPPP, by its inherent nature, behaves like a 1D PPP as $\theta \to 0$ and a 2D PPP as $\theta \to \infty$. By Theorems~\ref{th_zero} and \ref{th_infty}, the PLP-PPP exhibit the same behavior as the TPPP in the asymptotic regimes of $\theta$. Here, we analyze the non-asymptotic behavior of $M^{\mathrm{PLP-PPP}}_{1,m}$ and $M^\mathrm{TPPP}_{1,m}$. 
\begin{theorem}
The nearest-neighbor distance in the PLP-PPP is stochastically dominated by the distance from the typical vehicle at the origin in the TPPP to its nearest neighbor.
\label{th_nnf_sd}
\end{theorem}
\begin{IEEEproof}
Using $e^{-x} \geq 1-x$, we can bound the nearest-neighbor distance distribution in the PLP-PPP given in Lemma~\ref{nnf_plp_psp} as
\begin{align}
F^\mathrm{PLP-PPP}_{R}(r) & \leq 1- \exp(\hspace{-1mm} -\lambda  m r  - 4 \pi \mu \lambda  \smallint_{0}^{r} \hspace{-1mm}  \sqrt{r^2-u^2} \hspace{0.3mm}\mathrm{d}u ) \nonumber \\
& = 1- \exp(-\lambda  m r - \lambda  \mu \pi r^{2}), \nonumber \\
& \stackrel{(a)} = F_{R}^{\Psi_o^{m} \cup \Phi_2}(r) \stackrel{(b)}\equiv F_{R}^{\mathrm{TPPP}}(r),
\end{align}
where $(a)$ follows from the nearest-neighbor distance distributions of the 1D and 2D PPPs given in \eqref{eq:nnf_nd_ppp} with $\lambda_{1} = (m/2)\lambda $ and $\lambda_{2} = \lambda \mu$, and $(b)$ follows from Definition~\ref{def:TPPP}.
\end{IEEEproof}

\begin{conjecture}
\label{conj}
The distance from the typical vehicle at the origin to the $n$-th nearest neighbor in the TPPP stochastically dominates that in the PLP-PPP for all $n \in \mathbb{N}$. 
\end{conjecture}

An important consequence of Conjecture~\ref{conj} is that the success probability of the typical vehicle at the origin in the PLP-PPP is lower bounded by that in the TPPP. Here, we give a heuristic argument for Conjecture~\ref{conj}. The 2D density of the vehicles in the TPPP is the same as that in the PLP-PPP. Then the comparison of the distances to the $n$-th nearest neighbor $r_{n}$ can be based only on the vehicle placement with respect to the typical vehicle. 

\begin{figure}[]
\centering
\includegraphics[scale = 0.65] {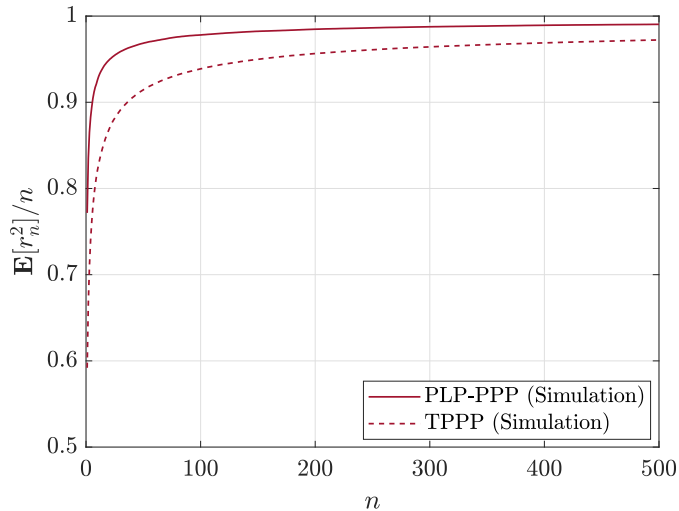}
 \caption{Comparison of normalized mean squared distances to the $n$-th nearest neighbor from the typical general vehicle at the origin in the PLP-PPP and TPPP. $\lambda = \mu = 1$.}
 \label{fig:mean_square_comp}
\end{figure}
Fig.~\ref{fig:mean_square_comp} compares the simulated values of $\mathbb{E}[r^{2}_{n}]/n$ in the PLP-PPP and TPPP. We observe that the mean squared distance from the typical general vehicle to the $n$-th nearest neighbor is higher for the PLP-PPP than the TPPP. The case of $n = 1$ follows from Theorem~\ref{th_nnf_sd}. We presume that this observation can be extended to higher values of $n$. Since the TPPP includes points at random independent locations compared to the PLP-PPP with points only concentrated on the lines, the probability that the $n$-th nearest neighbor is at a distance $r_{n}$ is higher for the TPPP. It follows that the average distance to the $n$-th nearest interferer from the origin is higher for the PLP-PPP, which in turn, leads to a higher success probability than for the TPPP.

\begin{figure}[]
\centering
\includegraphics[scale =0.65]{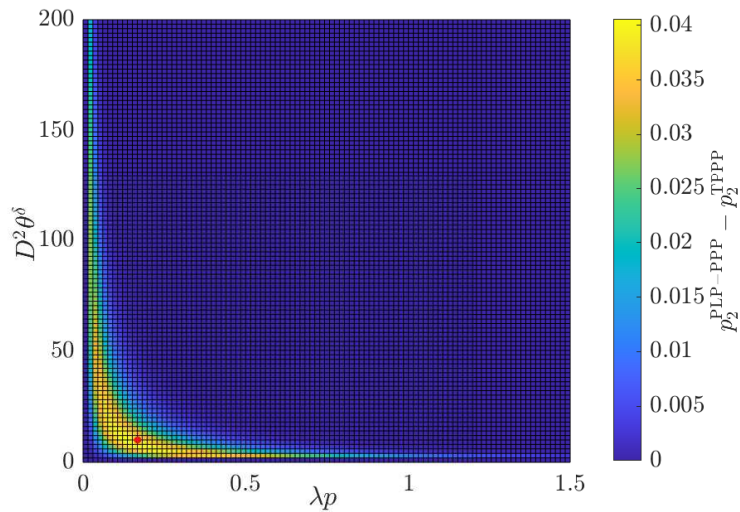}
\caption{Difference between the success probabilities of the typical general vehicle in the PLP-PPP~\eqref{eq:ps_plp_ppp} and TPPP~\eqref{eq:m1_tppp} as a function of $\lambda p$ and $D^2\theta^\delta$. $\mu = 0.204$, $\delta = 2/\alpha$, and $\alpha=4$. The maximum difference of $0.0404$ corresponding to the pair $(0.12, 10.1)$ is highlighted using a red filled circle. }
\label{fig:ps_plp_TPPP_comp2}
\end{figure}

Fig.~\ref{fig:ps_plp_TPPP_comp2} plots the difference in the success probabilities of the typical general vehicle in the PLP-PPP and TPPP. Letting $x=\lambda p$ and $y=D^2\theta^\delta$, the integrated difference $\int_0^\infty\int_0^\infty (p_2^{\mathrm{PLP-PPP}}(x,y)-p_2^{\mathrm{TPPP}}(x,y)) \hspace{0.2mm} \mathrm{d}x \hspace{0.2mm} \mathrm{d}y$ is maximized at $\mu=0.204$, which is the value we choose to plot the difference in Fig.~\ref{fig:ps_plp_TPPP_comp2}. The maximum difference between the success probabilities of the PLP-PPP and TPPP is about $-14$ dB ($0.0404$). Therefore, the success probability in the TPPP is a tight lower bound to that in the PLP-PPP. Note that the inferences obtained from Figs. \ref{fig:mean_square_comp} and~\ref{fig:ps_plp_TPPP_comp2} also apply to the typical intersection vehicle as the characterization of the streets that pass through the typical vehicle is the same in both the PLP-PPP and TPPP. 

\begin{figure}
\centering
\includegraphics[scale=.65]{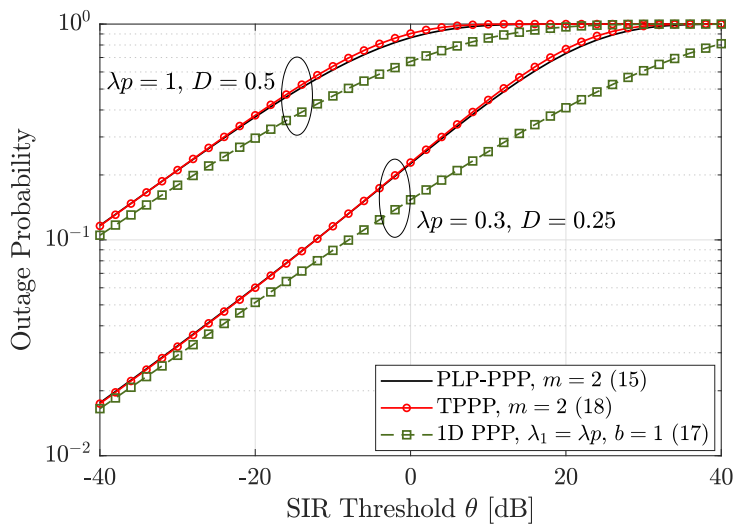}\label{fig:ps_plp_TPPP_comp1}
\caption{\label{fig:ps_plp_TPPP_comp} {Outage probabilities of the typical general vehicle in the PLP-PPP, TPPP, and that of the typical vehicle in a 1D PPP. $\mu = 1$ and $\alpha = 4$. The equation numbers corresponding to the success probability are given in the parentheses in the legend.}}
\end{figure}

Fig.~\ref{fig:ps_plp_TPPP_comp} compares the outage probabilities of the typical general vehicle in the PLP-PPP and TPPP to that of the typical vehicle in a 1D PPP. We observe that the TPPP better approximates the PLP-PPP for small $\theta$ than just a 1D PPP. As $\theta \to 0$, for SIR $> \theta$, there should not be any interferers in a small disk $b(o,r)$ of some radius $r$ centered at the origin. The pair correlation function for the PLP-PPP~\eqref{eq:pcf_plp} diverges as $r \to 0$, which indicates there definitely exists at least one line with vehicles of intensity $\lambda$ intersecting $b(o,r)$. For infinitesimally small $\theta$, only the typical street intersects $b(o,r)$. However, for non-vanishing values of $\theta$, there may be more than one line intersecting $b(o,r)$, and the 1D PPP is not sufficient to capture the effect of the streets other than the typical street intersecting $b(o,r)$.  

\begin{remark}
The success probability of the typical vehicle in the PLP-PPP can be tightly approximated by that in the TPPP. In particular, the approximations are asymptotically exact at the upper and lower tails of the success probability. 
\end{remark}
Next, we compare the moments of order $b > 1$ in the PLP-PPP and TPPP followed by their respective SIR meta distributions.

\subsection{Comparison of Higher-Order Moments and SIR Meta Distributions}
\begin{figure*}
\centering
\subfloat[$\theta = -20$ dB]{\includegraphics[scale = 0.4] {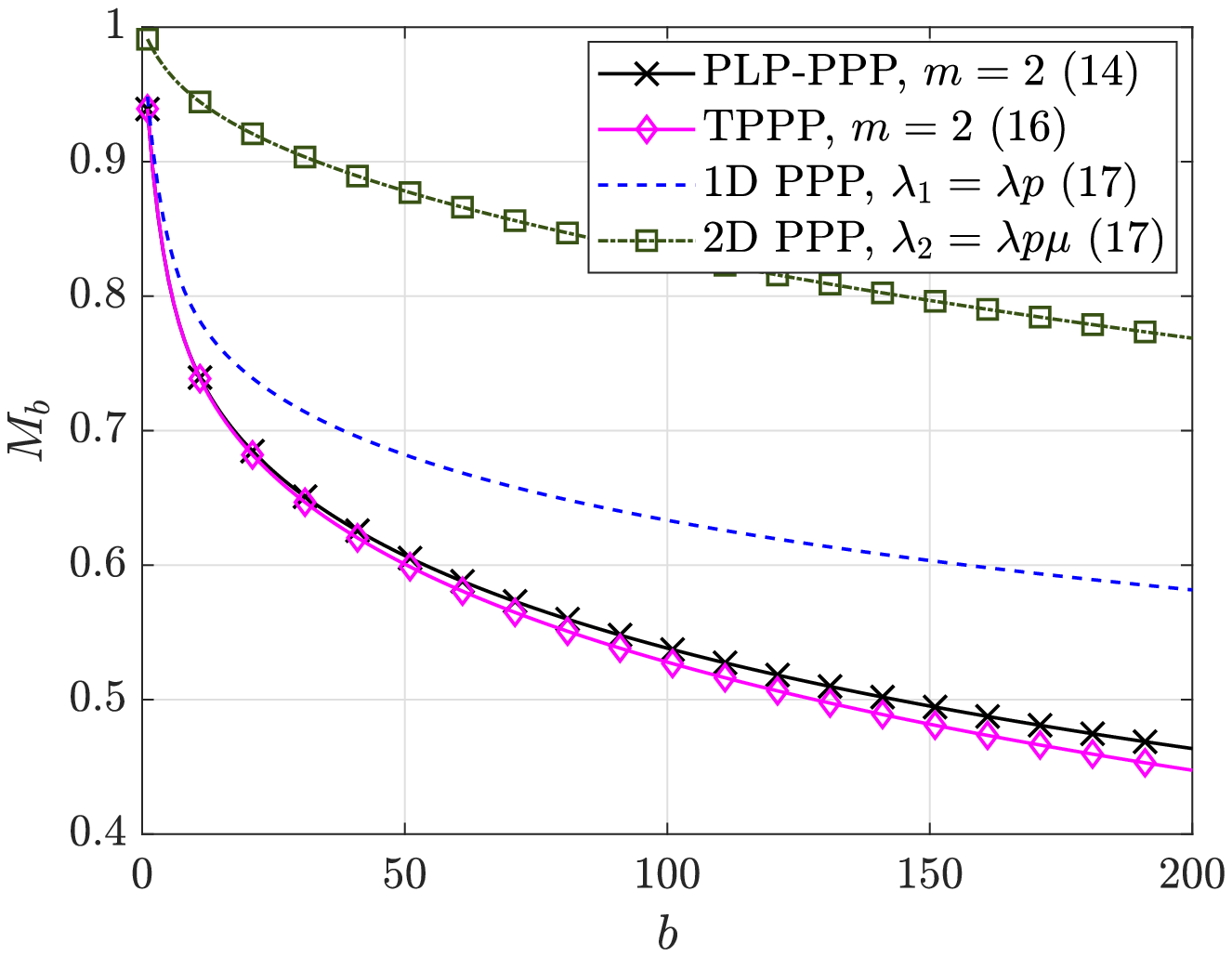}{\label{fig:mb_a}}
 }\hfill
\subfloat[$\theta = 0$ dB]{\includegraphics[scale = 0.4]{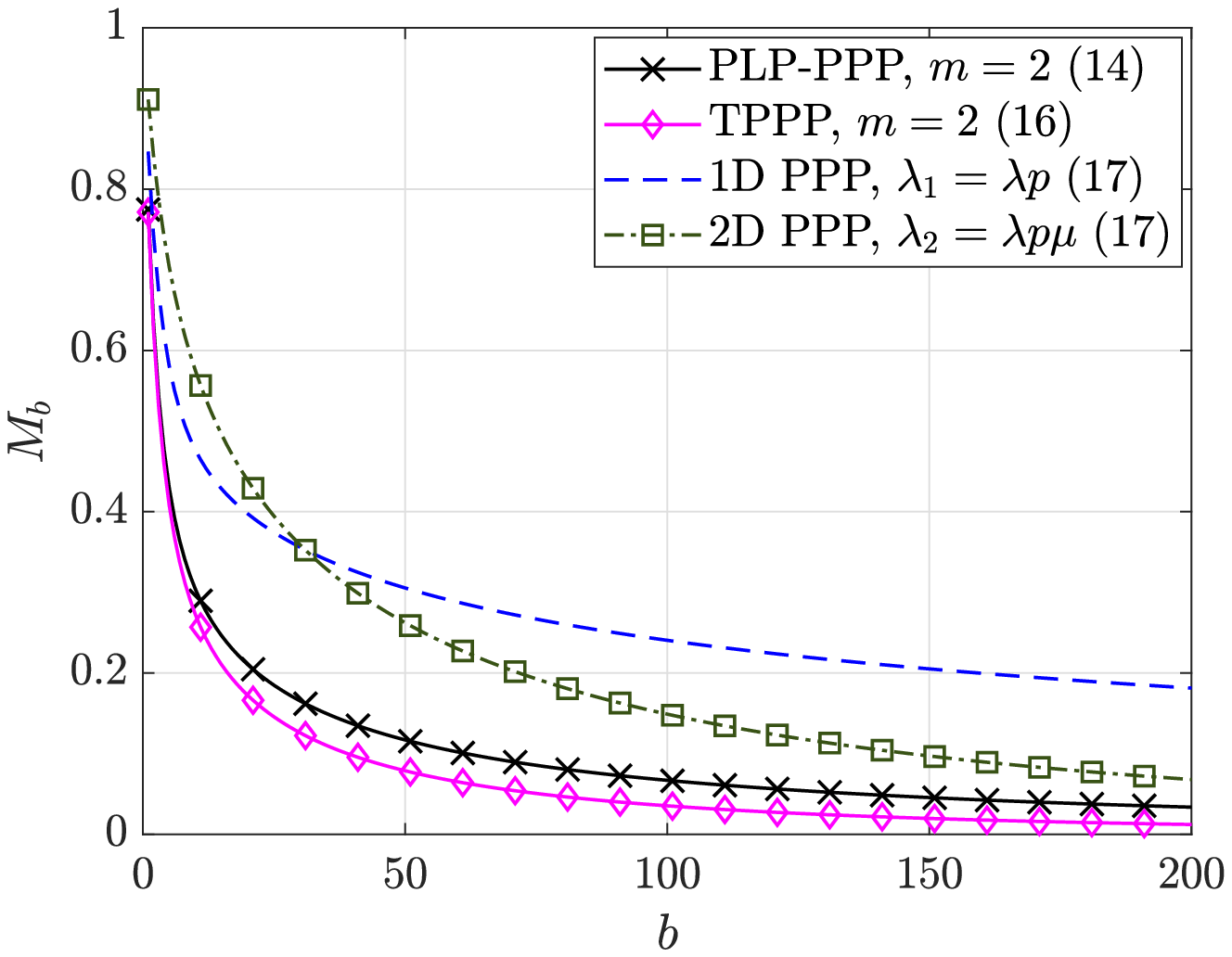}{\label{fig:mb_b}}
 }\hfill
\subfloat[$\theta = 10$ dB]{\includegraphics[scale = 0.4]{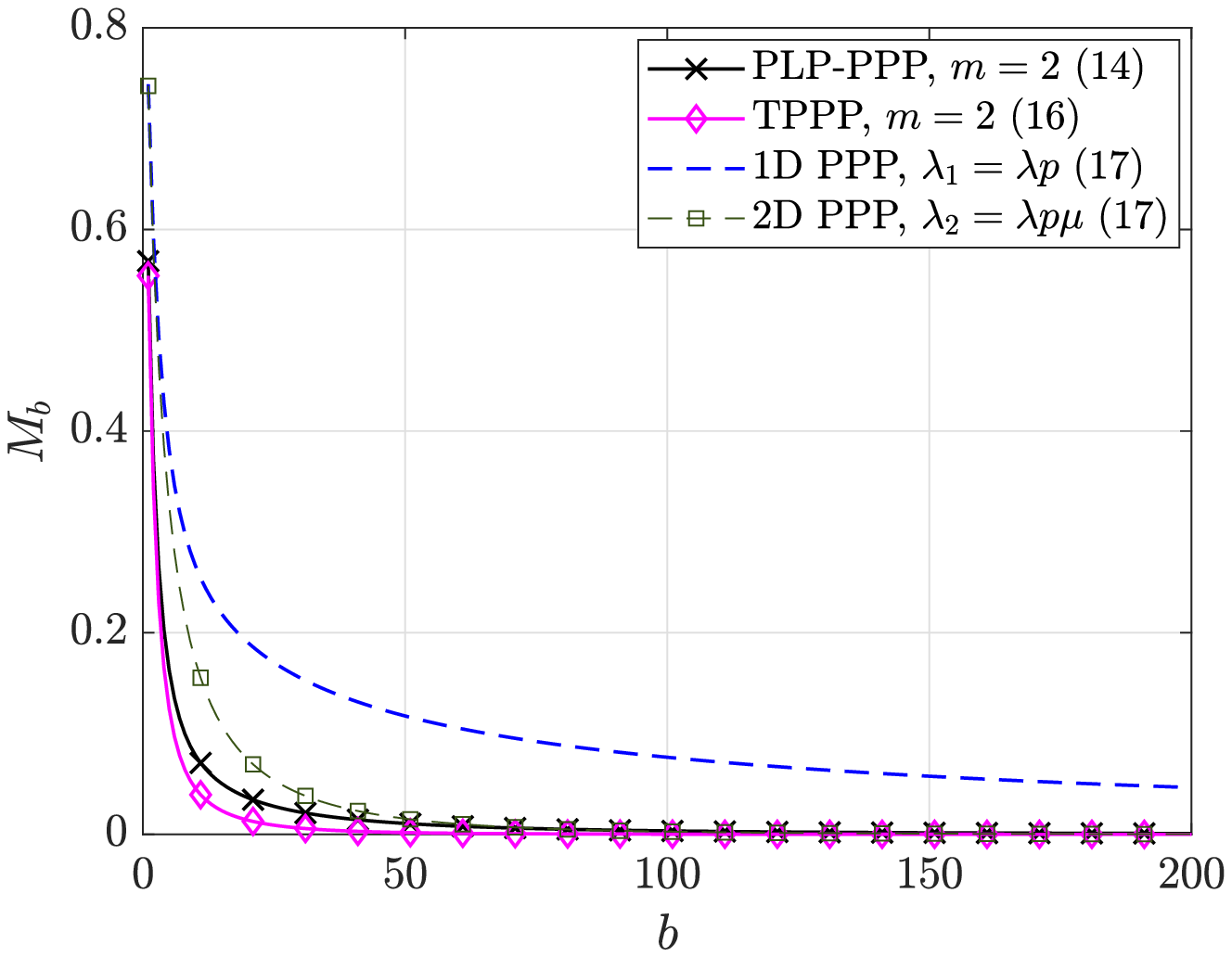}{\label{fig:mb_c}}
 }%
\caption{\label{fig:mb_comp} Moments of the conditional success probabilities for different SIR thresholds. $\mu = 1$, $\lambda = 1$, $p =0.3$, $D = 0.25$, and $\alpha = 4$. The equation numbers of the moments $M_b$ are given in the  parentheses in the legends.}
\end{figure*}

Fig.~\ref{fig:mb_comp} compares the moments of $P_{m}$ in the PLP-PPP, TPPP, 1D PPP, and 2D PPP. We observe that the moments in the PLP-PPP are lower bounded by that in the TPPP. Through Conjecture~\ref{conj}, we heuristically showed that the success probability of the typical vehicle in the PLP-PPP is lower bounded by that in the TPPP. The argument based on the stochastic dominance of the nearest-neighbor distance in the TPPP to the PLP-PPP in Conjecture~\ref{conj} extends to the moments of order $ b > 1$ as well. Also, we observe that the difference between the moments in the PLP-PPP and 2D PPP decreases with SIR threshold as in the case for $M_1(\theta)$ established in Theorem~\ref{th_infty}. 

\begin{figure*}
\centering
\subfloat[$x = 0.6$]{\includegraphics[scale = 0.55] {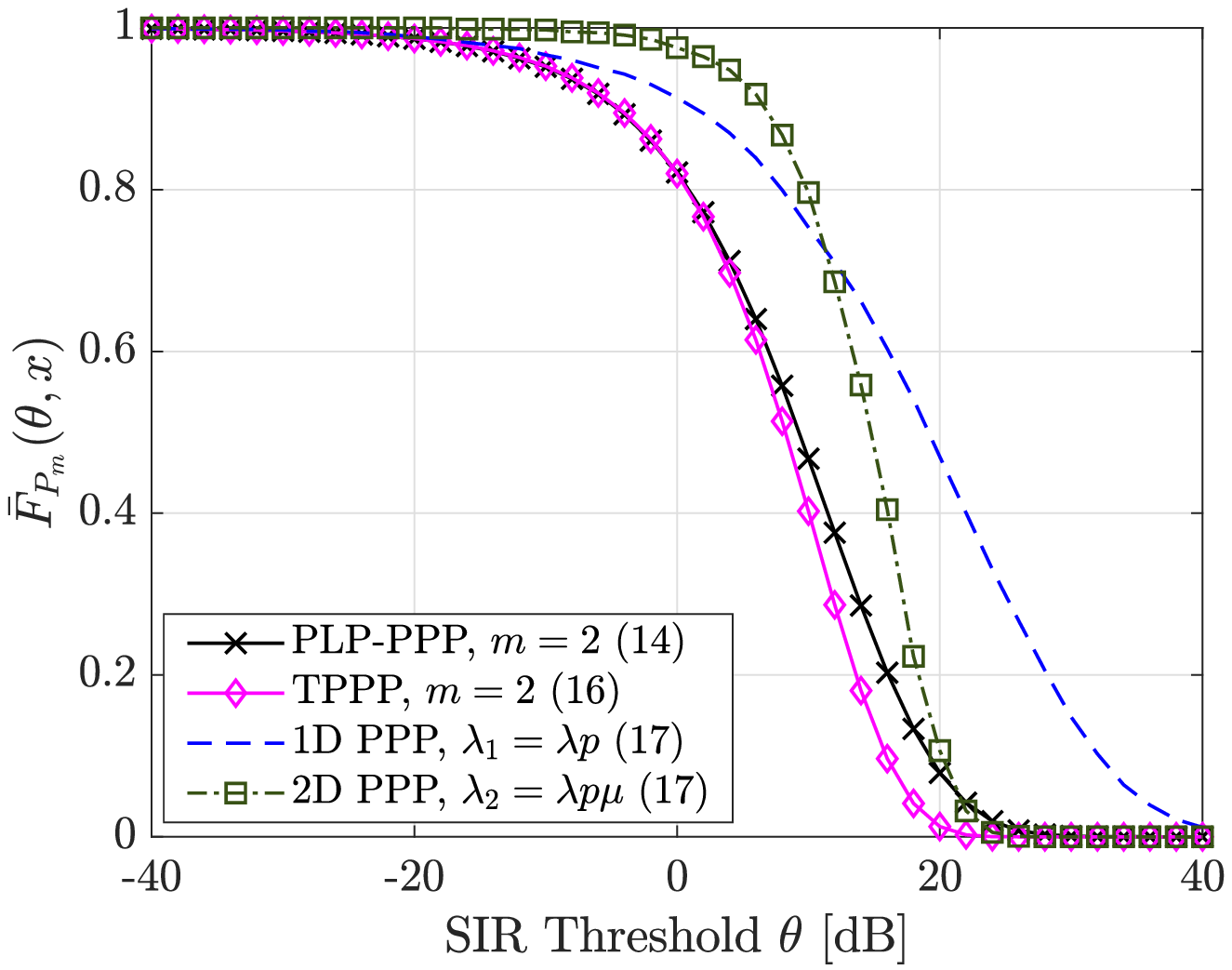}{\label{fig:mdx_a}}
 }\hfill
\subfloat[$x = 0.95$]{\includegraphics[scale = 0.55]{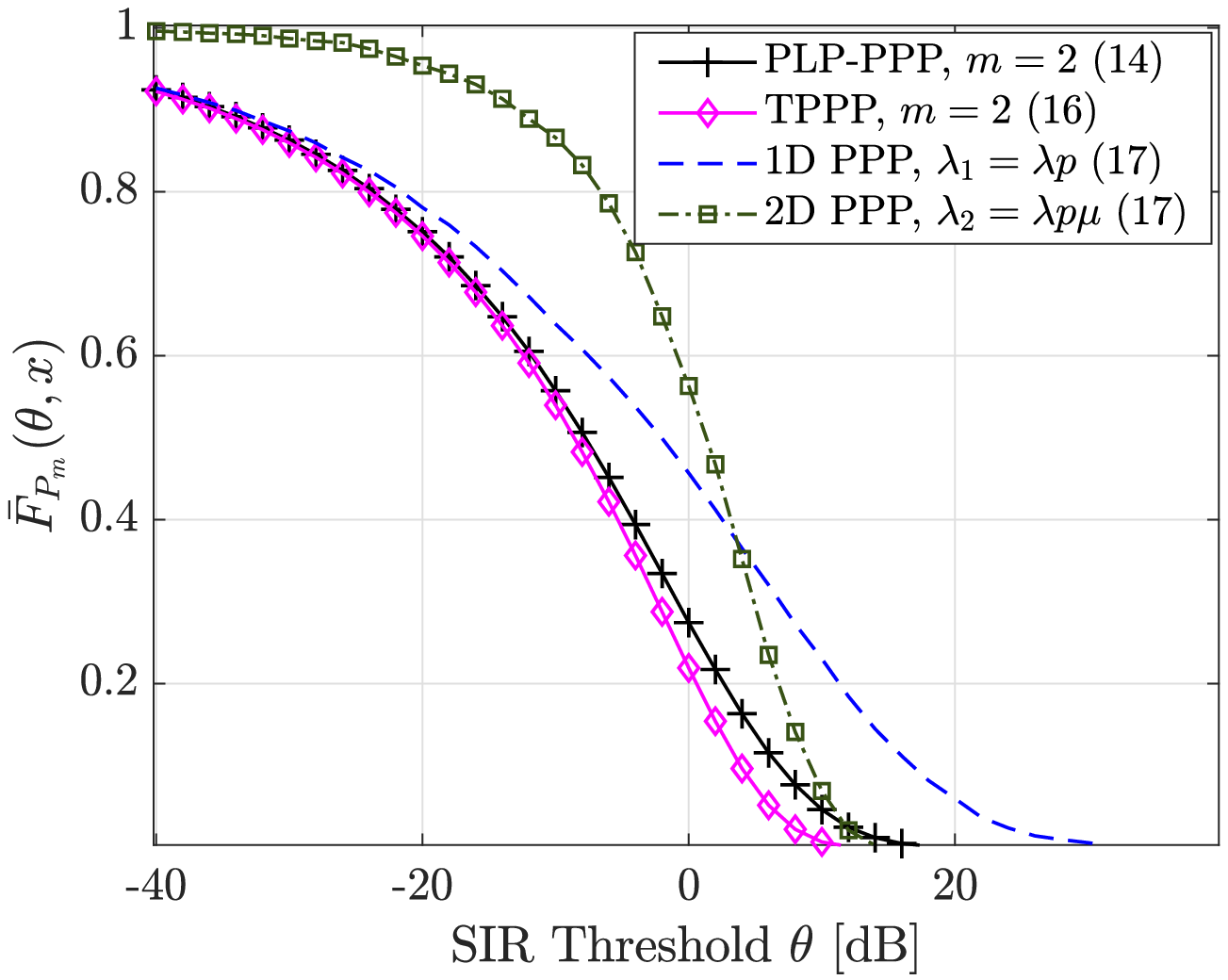}{\label{fig:mdx_b}}
 }%
\caption{\label{fig:md_given_x} SIR meta distributions for different reliabilities.  $\mu = 1$, $\lambda = 1$, $p =0.3$, and $D = 0.25$. The equation numbers of the moments required to evaluate $\bar{F}_{P_{2}}(\theta, x)$ are given in the parentheses in the legends.}
\end{figure*}

Fig.~\ref{fig:md_given_x} plots the SIR meta distributions for the PLP-PPP, TPPP, 1D PPP, and 2D PPP. For an SIR threshold $\theta$ of $0$ dB, $80\%$ of the links are at least $60\%$ reliable, whereas only $20\%$ of the links are at least $95\%$ reliable. To make $80\%$ of the links at least $95\%$ reliable, we need to reduce $\theta$ to $-24$ dB. Using the SIR MD, we can obtain the trade-offs between data rate (parametrized by $\theta$) and reliability. Also, we can find how to change the transmit probability $p$ to maintain a certain value of the MD, which we will discuss in detail in Section~\ref{sec:cc}. In terms of comparison with the TPPP, we observe that the SIR MDs for the TPPP and PLP-PPP are asymptotically exact with the gap being slightly larger in the middle ranges of $\theta$ than observed between their success probabilities (Fig.~\ref{fig:ps_plp_TPPP_comp2}). As the exact expression~\eqref{eq:exact_meta} also involves the moments of order $b > 1$, the differences between the moments $M^{\mathrm{PLP-PPP}}_{b,m}$ and $M^\mathrm{TPPP}_{b,m}$ combined produces a slightly larger gap than for the success probability (first moment). The PLP-PPP behaves like a 1D PPP as $\theta \to 0$ and 2D PPP as $\theta \to \infty$ as given in Theorems~\ref{th_zero} and~\ref{th_infty}. 

\subsection{Presence of Shadowing}
Now, let us assume that the channels are also subject to shadowing in addition to Rayleigh fading in the PLP-PPP and TPPP. Using~\eqref{eq:sir_eqn}, the SIR expression including shadowing can be written as
\begin{equation}
{\sf SIR} = \frac{g \nu D^{-\alpha}}{\sum_{z \in \chi} g_{z} \nu_{z} \Vert z \Vert^{-\alpha} B_{z}},
\label{eq:sir_eqn_shadow}
\end{equation}
where $\nu, \nu_{z}$ are i.i.d. shadowing random variables with mean $1$ and variance $\sigma^{2}$. $\chi = \mathcal{V}$ in the PLP-PPP, while $\chi = \mathcal{T}$ in the TPPP. 

\begin{theorem}
\label{th_shad}
Fix $\lambda' > 0$ and let the density of vehicles on each street be $\lambda=\lambda'/\mathbb{E}[\nu^\delta]$. Then, as $\sigma \to \infty$,
\[ \bar F_{P_m}^{\mathrm{PLP-PPP}}(\theta,x) \sim \bar F_{P_m}^{\mathrm{TPPP}}(\theta,x),\quad \theta\in\mathbb{R}^+, \hspace{1mm} x\in [0,1].\]
\end{theorem}
\begin{IEEEproof}
The PLP-PPP and TPPP differ only in the distribution of the vehicles that do not lie on the typical vehicle's streets.
Hence we need to show that the interference distributions from the rest of the vehicles that form the point processes $\mathcal{V}^!$ in the PLP-PPP and $\Phi_2$ in the TPPP are identical as  $\sigma \to \infty$. 
To this end, we focus on the propagation loss processes $\Upsilon_\chi \triangleq \lbrace \Vert z \Vert^{\alpha} /\nu_{z}: z \in \chi \rbrace$ for $\chi=\Phi_2$ and
$\chi=\mathcal{V}^!$.
By~\cite[Lemma 1]{zhang}, $\Upsilon_{\Phi_2}$ is a PPP on $\mathbb{R}^+$ with intensity function $\lambda(r)=\lambda'\mu\pi\delta r^{\delta-1}$.
By~\cite[Theorem 7]{bartek}, $\Upsilon_{\mathcal{V}^!}$ converges in distribution to a PPP with the same intensity function as $\sigma\to\infty$.
\end{IEEEproof}

The scaling of the density by $\mathbb{E}[\nu^\delta]$ in Theorem~\ref{th_shad} is necessary since without it, the intensity function of the one-dimensional point processes
$\Upsilon_{\chi}$ would go to $0$ or approach $\infty$ as $\sigma$ increases. While the convergence result would still hold, it would be trivial since in both models,
there would either be no interference or infinite interference.

A simple approximation to the SIR MD is obtained by just using the first two moments of the conditional success probability. As it varies between $0$ and $1$, the beta distribution characterized by the first two moments is a natural choice. It is shown that the beta distribution can tightly approximate the SIR meta distribution in Poisson bipolar and cellular networks~\cite{md_main}. In the next subsection, we explore whether the beta approximation works for vehicular networks.  

\subsection{Beta Approximation of the SIR Meta Distribution}
The probability density function (pdf) of a beta distributed random variable $X$ with parameters $\alpha$ and $\beta$ is given by
\begin{equation}
f_{X}(x) = \frac{x^{\alpha-1}(1-x)^{\beta-1}}{B(\alpha, \beta)},
\end{equation}
where $B(a,b) = \frac{\Gamma(a) \Gamma(b)}{\Gamma(a+b)}$. The first and second order moments of $X$ are 
\begin{align}
\mathbb{E}[X] = \frac{\alpha}{\alpha+\beta}, \hspace{2mm} \text{and} \hspace{2mm} \mathbb{E}[X^{2}] =  \frac{\alpha+1}{\alpha+\beta+1}\mathbb{E}[X],
\end{align}
respectively. We obtain $\alpha$ and $\beta$ by equating $\mathbb{E}[X] = M_{1,m}(\theta)$ and $\mathbb{E}[X^{2}] = M_{2,m}(\theta)$. The complementary cumulative distribution function of $X$ is the beta approximation of the SIR meta distribution, {\em{i.e.,}}
\begin{align}
\bar{F}_{P_{m}}(x) \approx 1-I_{x}(\alpha, \beta),
\label{eq:beta_app}
\end{align}
where $I_{x}(\alpha, \beta)$ is the regularized incomplete beta function.

\begin{figure*}
\centering
\subfloat[]{\includegraphics[scale = 0.52] {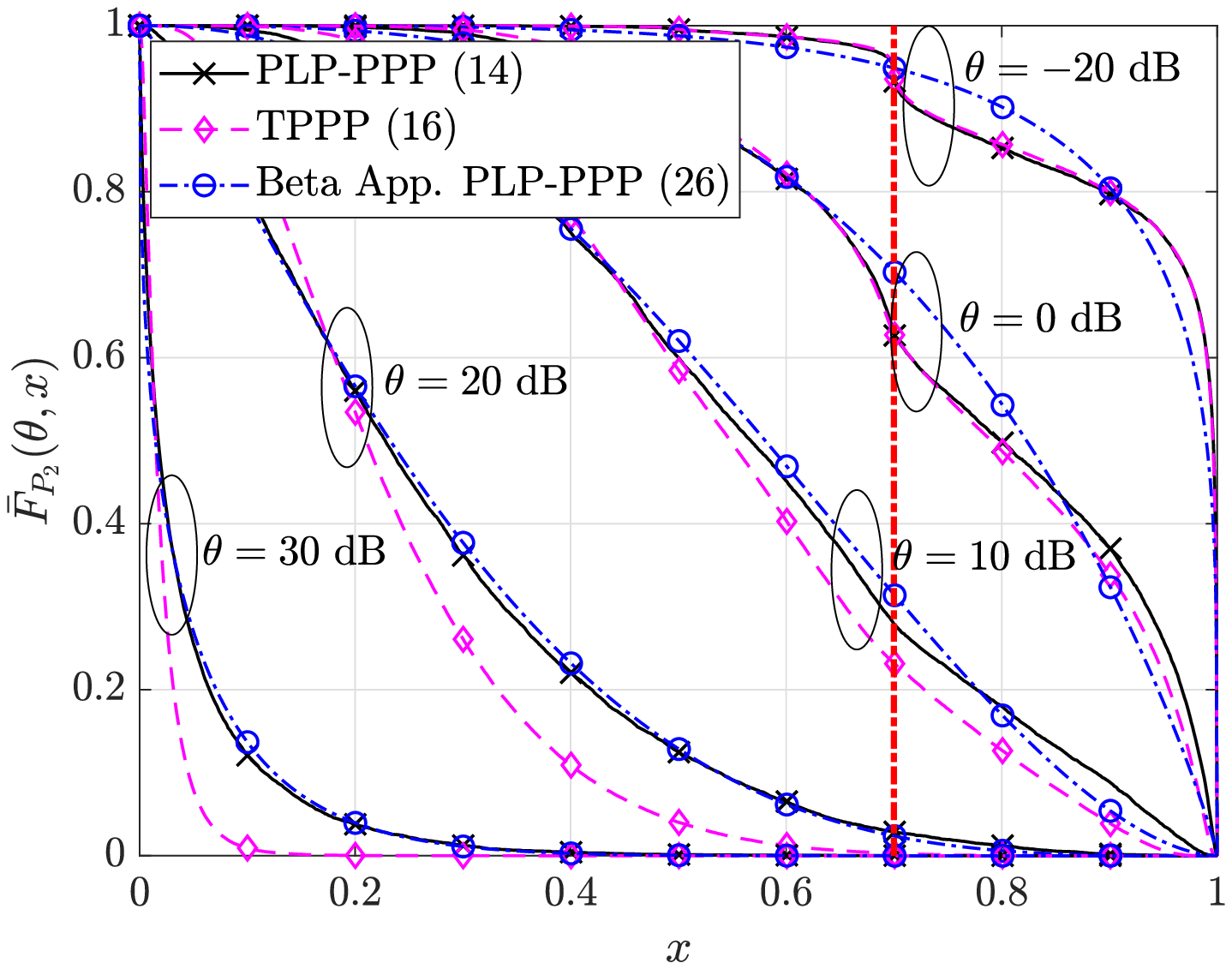}{\label{fig:exact_vs_beta_plp}}} \hspace{10mm}
\subfloat[]{\includegraphics[scale = 0.52]{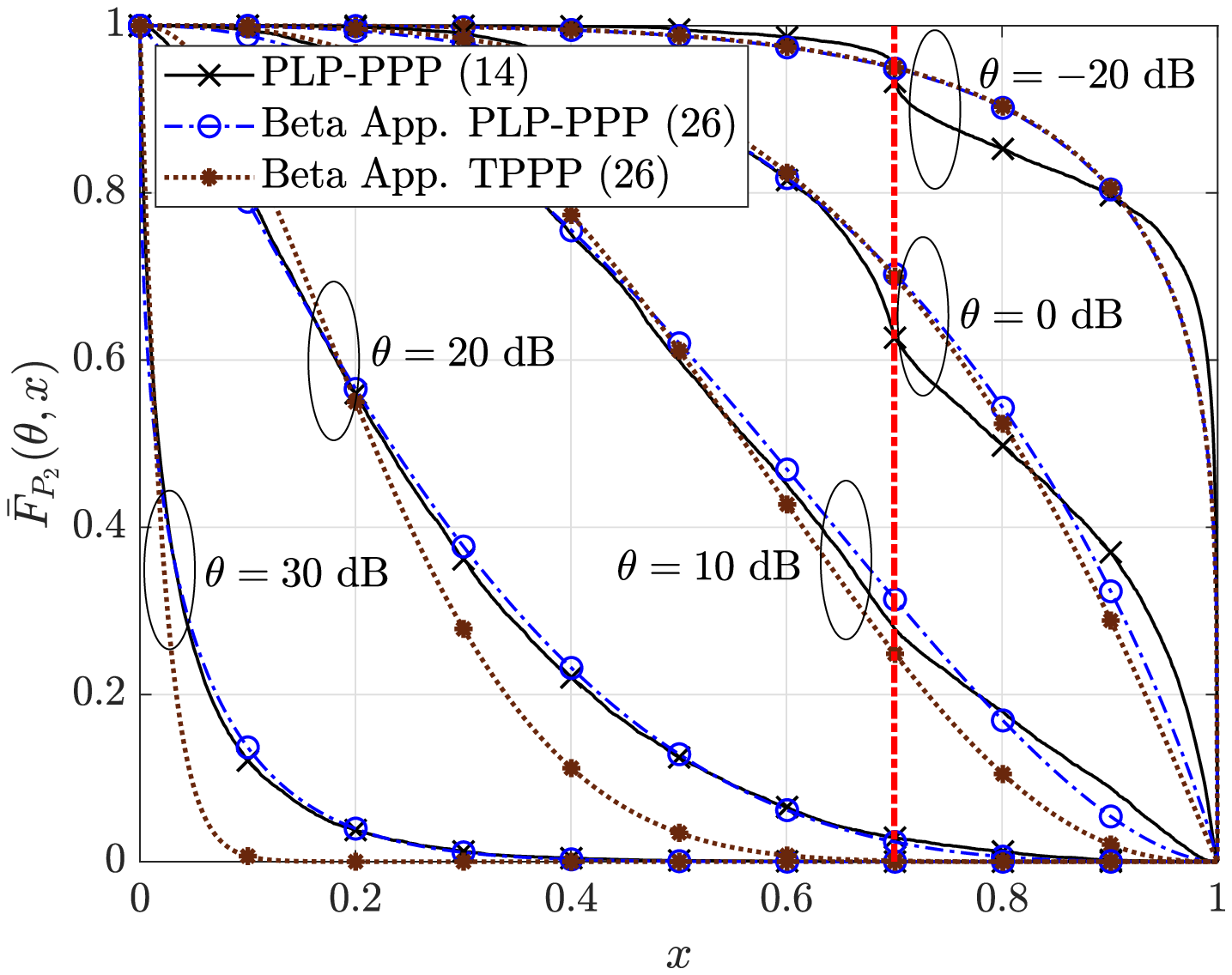}
{\label{fig:exact_vs_beta_tppp}}}%
\caption{\label{fig:exact_vs_beta} SIR meta distributions for the PLP-PPP, TPPP and their beta approximations. $\mu = 1$, $\lambda = 1$, $p =0.3$, $\alpha = 4$, and $D = 0.25$. The transition at $x = 1-p$ is highlighted by the dashed line. The equation numbers in the parentheses in the legends either refer to the moments or expression of $\bar{F}_{P_{2}}(\theta, x)$.}
\end{figure*}

Fig.~\ref{fig:exact_vs_beta} shows a different cross-section of the SIR MD of the typical general vehicle in the PLP-PPP and compares it to that in the TPPP and the beta approximations for the PLP-PPP and TPPP. The SIR MD for the PLP-PPP tightly approximates that for the TPPP in the asymptotic regimes of $\theta$ as in Fig.~\ref{fig:md_given_x}, and in that of $x$. At $x = 1-p$, which is $0.7$ in the considered network setting, there is a transition in the meta distribution curves, particularly noticeable at lower SIR thresholds. The reason is that as $\theta \to 0$, there should not be any interferers in a small disk around the typical vehicle. If the nearest interferer is absent with probability $1-p$, then there exists a non-zero fraction of links that can satisfy a reliability of $1-p$. We can neglect the case of two or more interferers present within that small distance to the typical vehicle, as the probability of such an event vanishes asymptotically. We observe that the beta approximation is tight at higher SIR thresholds, whereas at lower SIR thresholds, the first two moments that define the beta approximation are not sufficient to tightly characterize the transition at $x = 1-p$ in both the PLP-PPP (Fig.~\ref{fig:exact_vs_beta_plp}) and TPPP (Fig.~\ref{fig:exact_vs_beta_tppp}). Instead, the beta approximation smoothes out the meta distribution. Furthermore, we can doubly approximate the SIR MD for the PLP-PPP by the beta approximation of the SIR MD for the TPPP (Fig.~\ref{fig:exact_vs_beta_tppp}), which is tight in the asymptotic regimes of $\theta$ and $x$. 

\begin{remark}
\textit{
\begin{itemize}
\item[1.] The maximum difference between the success probabilities $M_{1,m}^{\mathrm{PLP-PPP}}$ and $M_{1,m}^{\mathrm{TPPP}}$ is about $-14$ dB over the entire parameter space (Fig.~\ref{fig:ps_plp_TPPP_comp2}).  
\item[2.] Fig.~\ref{fig:exact_vs_beta_plp} suggests that there exists some $\hat{\theta}$ such that the SIR MD for the PLP-PPP tends to its beta approximation for $\theta > \hat{\theta}$ and to that for the TPPP for $\theta \leq \hat{\theta}$, $\forall x$. Hence, it is not always necessary to evaluate all the $b-$th ($b \in \mathbb{N}$) moments of the conditional success probability to evaluate the SIR MD.
\item[3.] The beta approximation of the SIR MD for the TPPP is fairly accurate $\forall \theta$ and $\forall x$ and becomes increasingly tight as $\theta \to 0$ or $\infty$ and $x \to 0$ or $1$ (Fig.~\ref{fig:exact_vs_beta_tppp}). 
\item[4.] The accuracy of the TPPP further improves under shadowing. The SIR MD for the TPPP approaches that for the PLP-PPP as the variance of the shadowing increases. This implies that the worst-case TPPP approximation is the case of no shadowing. 
\end{itemize}}
\end{remark}

\section{The Transdimensional Approach to the PSP-PPP}
First, we analyze the first moment of the conditional success probability, and then the SIR MDs for the PSP-PPP and corresponding TPPP as in~Section~\ref{sec:ltppp}.
\subsection{First-Order Moments: PSP-PPP vs. TPPP}
By~\eqref{eq:mb_prod}, the moment $M_{b,m}$ can be expressed as $
M_{b,m} = M^{o}_{b,m} M^{!}_{b,m}$, where $M^{o}_{b,m}$ considers only the interference from the vehicles on the typical vehicle's streets, and $M^{!}_{b,m}$ takes into account the interference from the vehicles on the rest of the streets. 
\begin{lemma}[{{{\hspace{-0.1mm}\cite{jeyaj_equiv}, Prop. 2}}}]
\label{prop_lio_psp}
For the PSP-PPP with half-length density function $f_{H}(h)$, we have $M^{o}_{b,m}$ given by
\begin{align}
M^{o}_{1,m}(s) = & \bigg(\int \limits_{0}^{\infty} \bigg( \frac{1}{2h} \int \limits_{-h}^{h} \exp \bigg(- \lambda p  s^{\delta/2}  \int \limits_{(-w-h) s^{-\delta/2}}^{(-w+h) s^{-\delta/2}} \frac{1}{ 1 + v^{{2/\delta}}} \mathrm{d}v\bigg) \mathrm{d}w \bigg)\tilde{f}_{H}(h) \hspace{0.3mm} \mathrm{d}h \bigg)^{m/2}, \label{eq:lio_psp} 
\end{align}
 where $s = \theta D^{\alpha}$, $m \in \lbrace 2, 4 \rbrace$, and $\tilde{f}_{H}(h) = hf_{H}(h)/\mathbb{E}[H]$.
\end{lemma}

\begin{lemma}[{{{\hspace{-0.1mm}\cite{jeyaj_equiv}, Eqn. (18)}}}]
\label{prop_ps_psp}
The success probability $p_{m}^{\mathrm{PSP-PPP}}$, or the first moment of the conditional success probability $M_{1,m}^{\mathrm{PSP-PPP}}$ is given by
\begin{align}
p_{m}^{\mathrm{PSP-PPP}}  = & M^{o}_{1,m}(\theta D^{\alpha}) \exp\bigg(-\frac{\mu}{\pi} \int \limits_{0}^{\infty} \int \limits_{0}^{\pi} \int \limits_{0}^{2 \pi} \int \limits_{0}^{\infty} (1 - \mathcal{L}_{I_{a}}(\theta D^{\alpha})) \gamma {f}_{H}(h)  \hspace{0.3mm}\mathrm{d}\gamma \hspace{0.3mm} \mathrm{d}\phi \hspace{0.3mm} \mathrm{d}\varphi \hspace{0.3mm} \mathrm{d}h \bigg), \label{eq:ps_psp_ppp}
\end{align}
 where $M^{o}_{1,m}(\theta D^{\alpha})$ is given by Lemma~\ref{prop_lio_psp}, $\mathcal{L}_{I_{a}}(s) = \exp\big(-\lambda p \displaystyle \smallint_{-h}^{h} \big({1+ f(\gamma, \phi, \varphi, u)^{1/\delta}}\big)^{-1} \mathrm{d}u \big)$ with $f(\gamma, \phi, \varphi, u) = (\gamma^2+u^2+2 \gamma u \cos(\phi-\varphi))s^{-\delta}$,  and  $m \in \lbrace 2, 4 \rbrace$.
\end{lemma}
The success probability~\eqref{eq:ps_psp_ppp} tends to that in a point process formed only on the typical vehicle's streets as $\theta \to 0$ and a 2D PPP as $\theta \to \infty$~\cite[Lemmas 8 and 9]{jeyaj_equiv}. Next, we derive the first moment for the transdimensional model of the PSP-PPP formed by the superposition of the point process on the typical vehicle's streets and a 2D PPP. 

\begin{figure*}[!t]
\centering
\subfloat[$\mu = 0.01$]{\includegraphics[scale = 0.52] {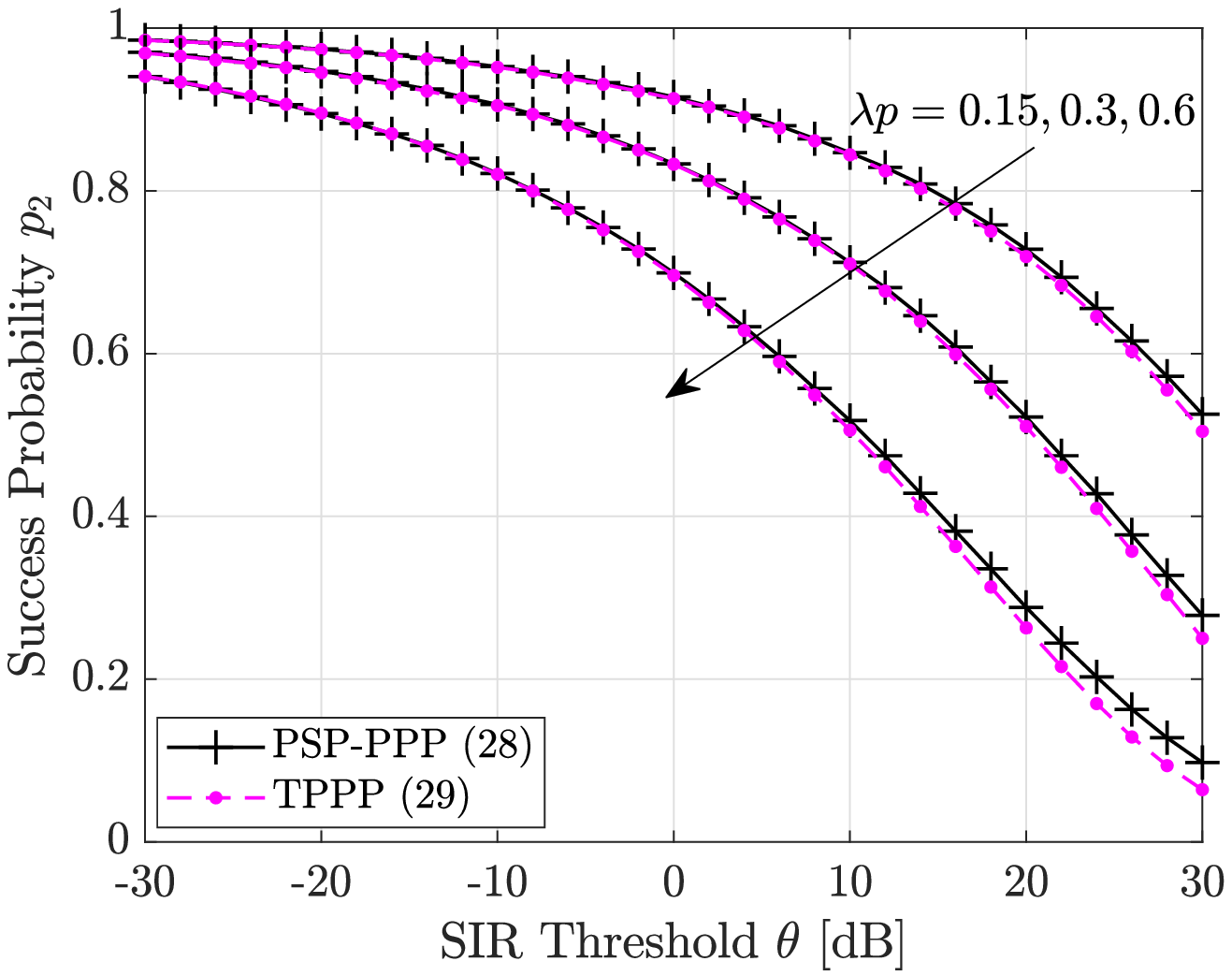}{\label{fig:stpp_a}}
 }\hspace{15mm}
\subfloat[$\mu = 1$]{\includegraphics[scale = 0.52]{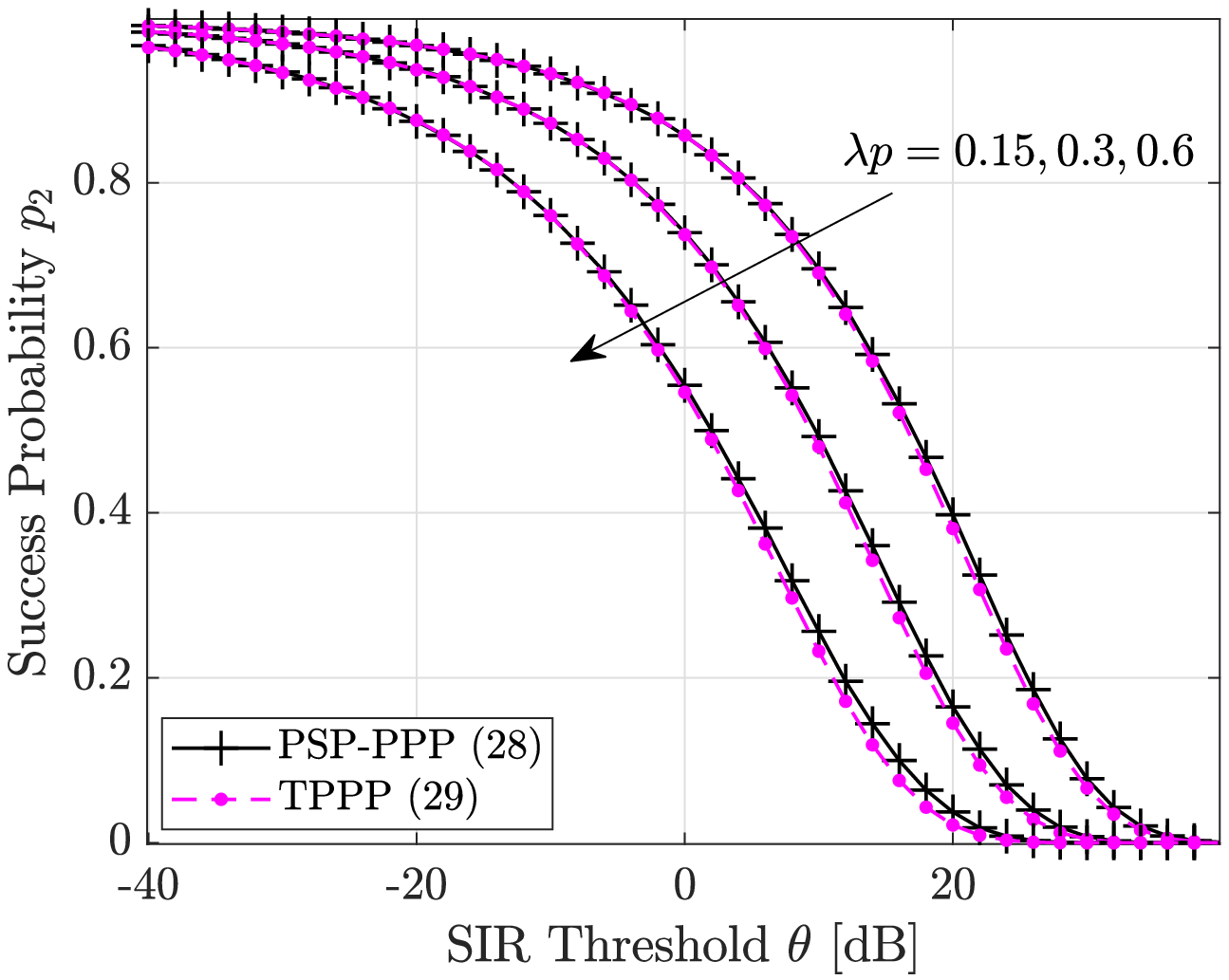}{\label{fig:stpp_b}}
}\hfill
\subfloat[$\mu = 0.01$]{\includegraphics[scale = 0.52]{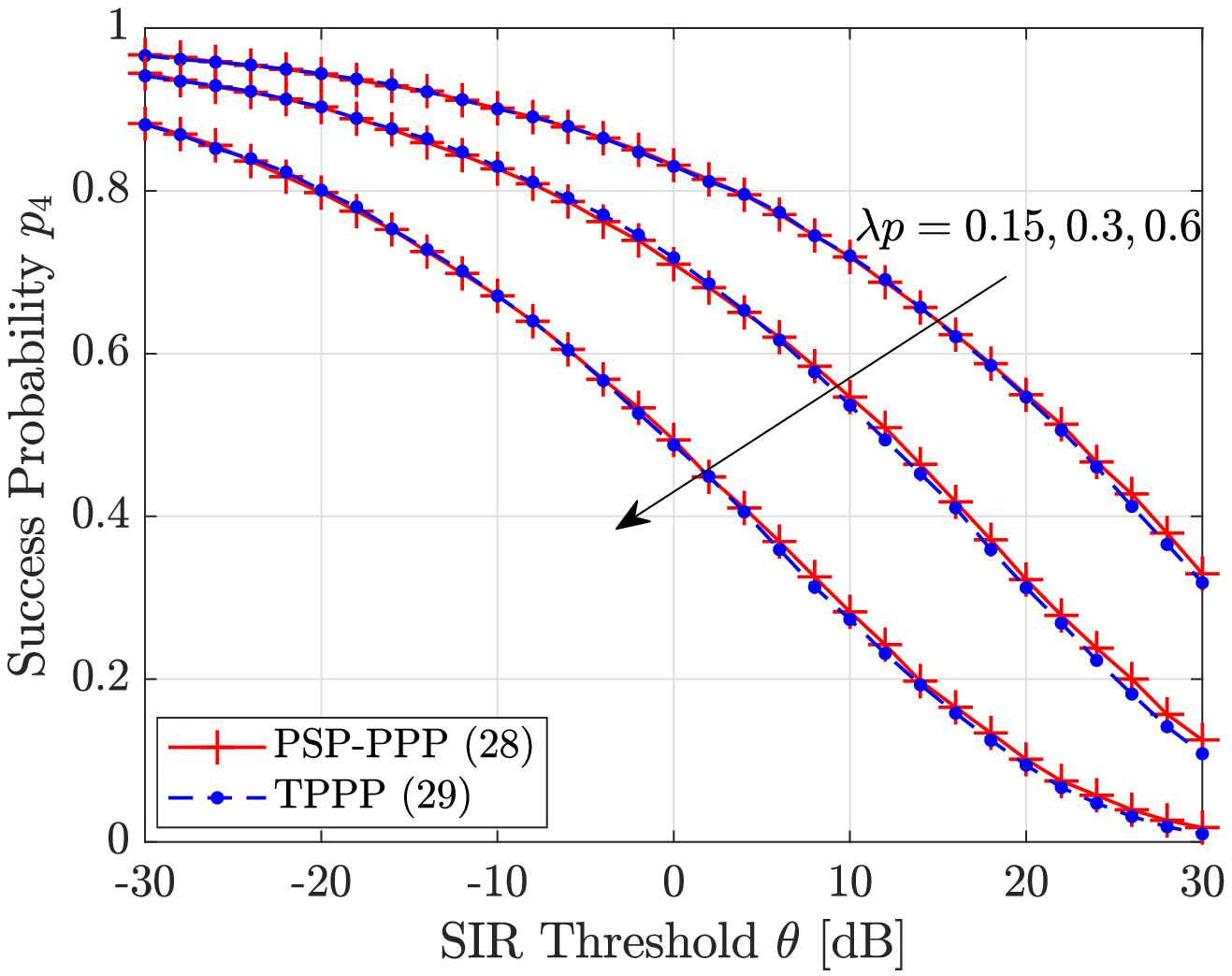}{\label{fig:stpp_c}}
 }\hspace{15mm}
\subfloat[$\mu = 1$]{\includegraphics[scale = 0.52]{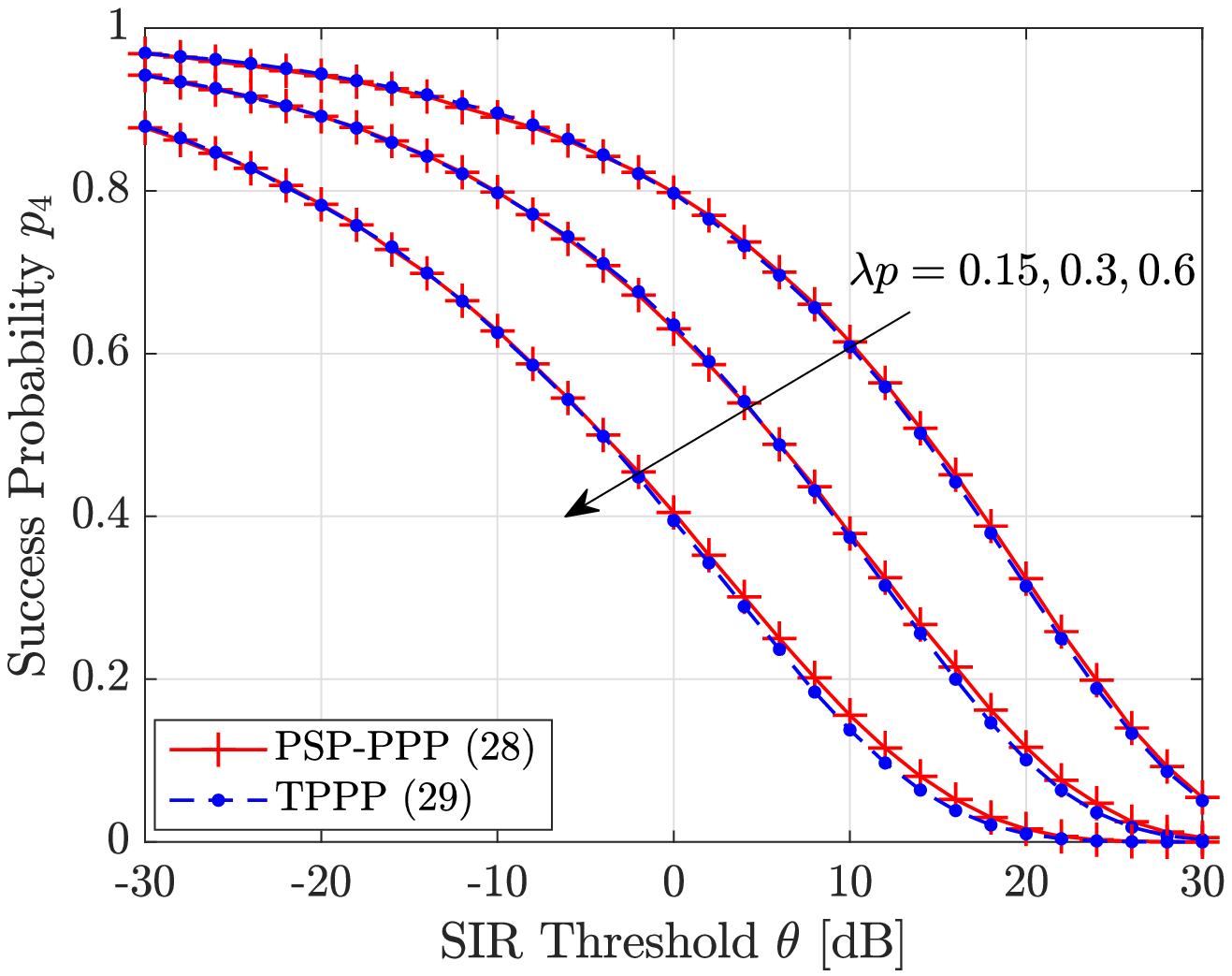}{\label{fig:stpp_d}}
 } %
\caption{\label{fig:stpp_comp} Success probabilities of the typical general ((a) and (b)) and intersection ((c) and (d)) vehicles in the PSP-PPP and the corresponding TPPP.  $f_{H}(h) = 2ch\exp(-ch^{2})$ with $c = \mu$, $D = 0.25$, and $\alpha = 4$. }
\end{figure*}

\begin{lemma}
The success probability of the typical vehicle of order $m \in \lbrace 2, 4 \rbrace$ at the origin in the TPPP corresponding to the PSP-PPP is given by
\begin{align}
p_{m}^{\mathrm{TPPP}} = M^{o}_{1,m}(\theta D^{\alpha}) \exp( -& 2\lambda p \mu \pi \mathbb{E}[H] D^{2} \theta^{\delta}   \Gamma(1+\delta)\Gamma(1-\delta)),
\label{eq:ps_stppp}
\end{align}
where $M^{o}_{1,m}(\theta D^{\alpha})$ is given by~\eqref{eq:lio_psp}, and $\delta = 2/\alpha$. 
\end{lemma}
\begin{IEEEproof}
The point processes $\Psi_o^{m}$ and $\Phi_2$ forming the TPPP are independent. It follows that the success probability of the typical vehicle at the origin in the TPPP corresponding to the PSP-PPP is the product of $M^{o}_{1,m}(\theta D^{\alpha})$ given by~\eqref{eq:lio_psp} and the success probability of the typical vehicle in $\Phi_2$ given by~\eqref{eq:mb_nd_ppp} with $b = 1$. The intensity of active transmitters in $\Phi_2$ is $2\lambda p \mu \mathbb{E}[H]$ by Lemma~\ref{lemma_lsp_2d}.
\end{IEEEproof}

Fig.~\ref{fig:stpp_comp} compares the success probabilities of the typical vehicle in the PSP-PPP and the corresponding TPPP for different values of $\lambda$ and $\mu$. We observe that the success probability of the typical vehicle in the TPPP tightly lower bounds that in the PSP-PPP. The reason is that the probability of finding the $n$-th nearest neighbor within a distance $r$ is higher in the TPPP than in the PSP-PPP as the vehicles are randomly placed on the plane without clustering to the streets. We presume that Conjecture~\ref{conj} that focuses on the stochastic dominance of the $n$-th nearest neighbor holds for the PSP-PPP as well. The case of $n=1$ can be proved  similarly to Theorem~\ref{th_nnf_sd} using \eqref{eq:nnf_nd_ppp} and~\eqref{eq:nnf_psp}.
\begin{figure}
\centering
{\includegraphics[scale = 0.62]{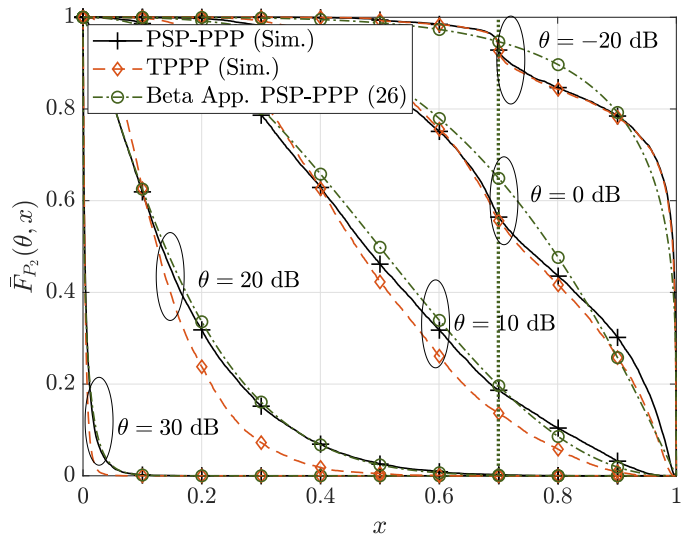}}
\caption{\label{fig:exact_vs_beta_psp} Comparison of exact SIR meta distribution for the PSP-PPP and its approximations.  $f_{H}(h) = 2ch\exp(-ch^{2})$ with $c=1$, $\mu = 1$, $\lambda = 1$, $p =0.3$, $\alpha = 4$, and $D = 0.25$. The transition at $x = 1-p$ is highlighted by the dotted line. The beta approximation to $\bar{F}_{P_{2}}(\theta, x)$ is given by~\eqref{eq:beta_app}.}
\end{figure}
\subsection{SIR Meta Distribution: PSP-PPP vs. TPPP}
We observe from Lemmas~\ref{prop_lio_psp} and \ref{prop_ps_psp} that the first moment of the conditional success probability for the PSP-PPP involves multiple nested integrals. The moments of order $b > 1$ are even more complicated and hence omitted. We analyze the SIR meta distributions for the PSP-PPP through simulations. Fig.~\ref{fig:exact_vs_beta_psp} compares the SIR meta distributions for the PSP-PPP, the TPPP, and the beta approximation. The behavior is the same as observed in Fig.~\ref{fig:exact_vs_beta} for the PLP-PPP, and thus Remark 2 also holds for the PSP-PPP. 

\begin{remark}
The success probability expression of the PSP-PPP~\eqref{eq:ps_psp_ppp} involving multiple integrals can be approximated by a much simpler expression~\eqref{eq:ps_stppp} obtained by its transdimensional model. In the case of the SIR meta distribution, the transdimensional model well approximates the PSP-PPP, especially, in the asymptotic regimes of $\theta$ and $x$ (Fig.~\ref{fig:exact_vs_beta_psp}). 
\end{remark}

\section{Application to Congestion Control}
\label{sec:cc}
We have established that the TPPP is sufficient to analyze the complicated PLP-PPP and PSP-PPP. Particularly, the beta approximation of the SIR MD for the TPPP provides tight approximations to the SIR MD for the PLP-PPP and PSP-PPP in the asymptotic regimes of $x$ and $\theta$. In this section, we introduce the transmit rate control in the PLP-PPP using the beta approximation of the SIR MD of the TPPP. The insights presented in this section also hold for the PSP-PPP. First, we begin with success probability-based congestion control to demonstrate the need for SIR MD-based congestion control. 

\subsection{Success Probability-Based Congestion Control} 

\begin{figure*}[!t]
\centering
\subfloat[]{\includegraphics[scale = 0.55]{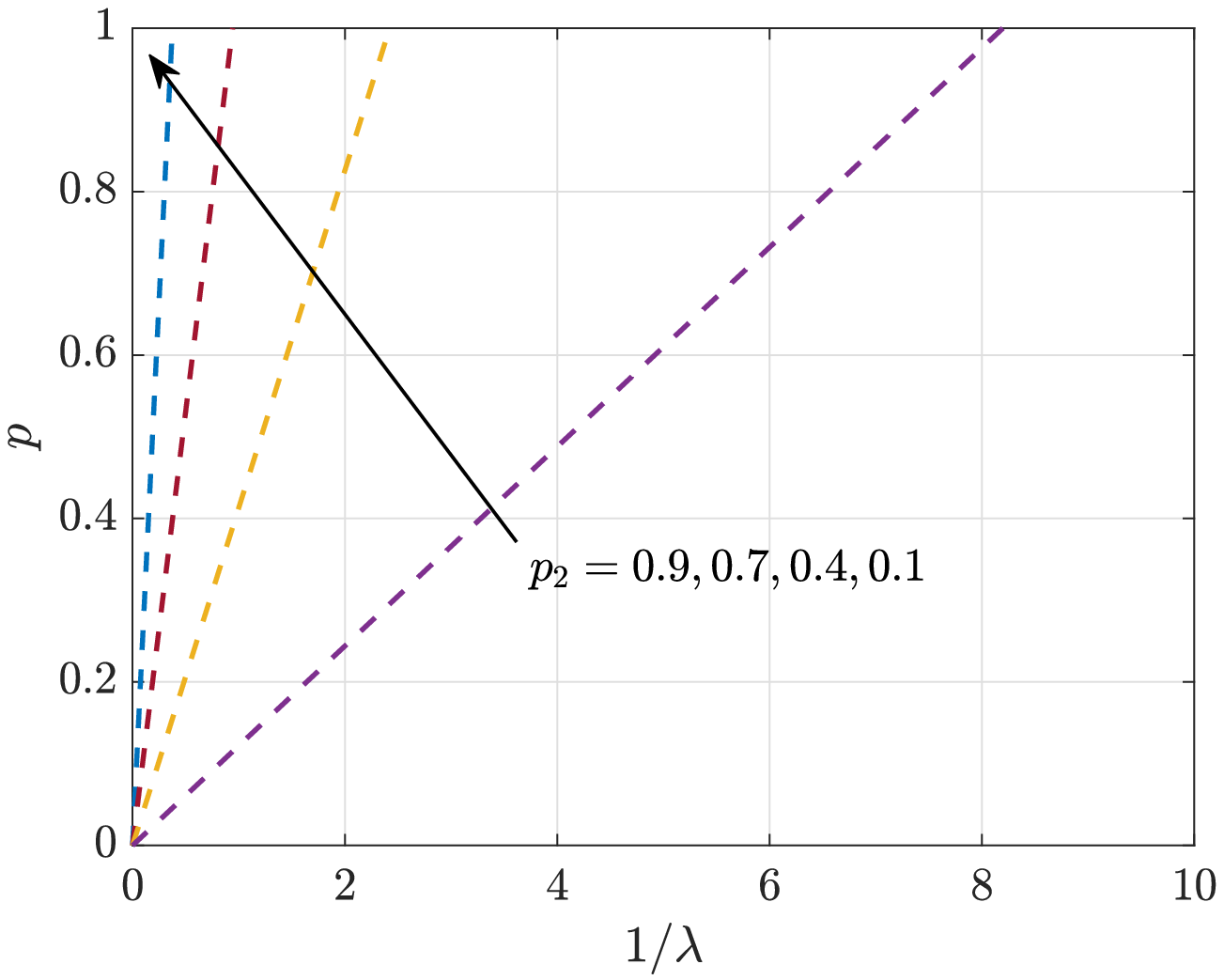}{\label{fig:mean_cc_a}}} \hspace{10mm}
\subfloat[]{\includegraphics[scale = 0.55]{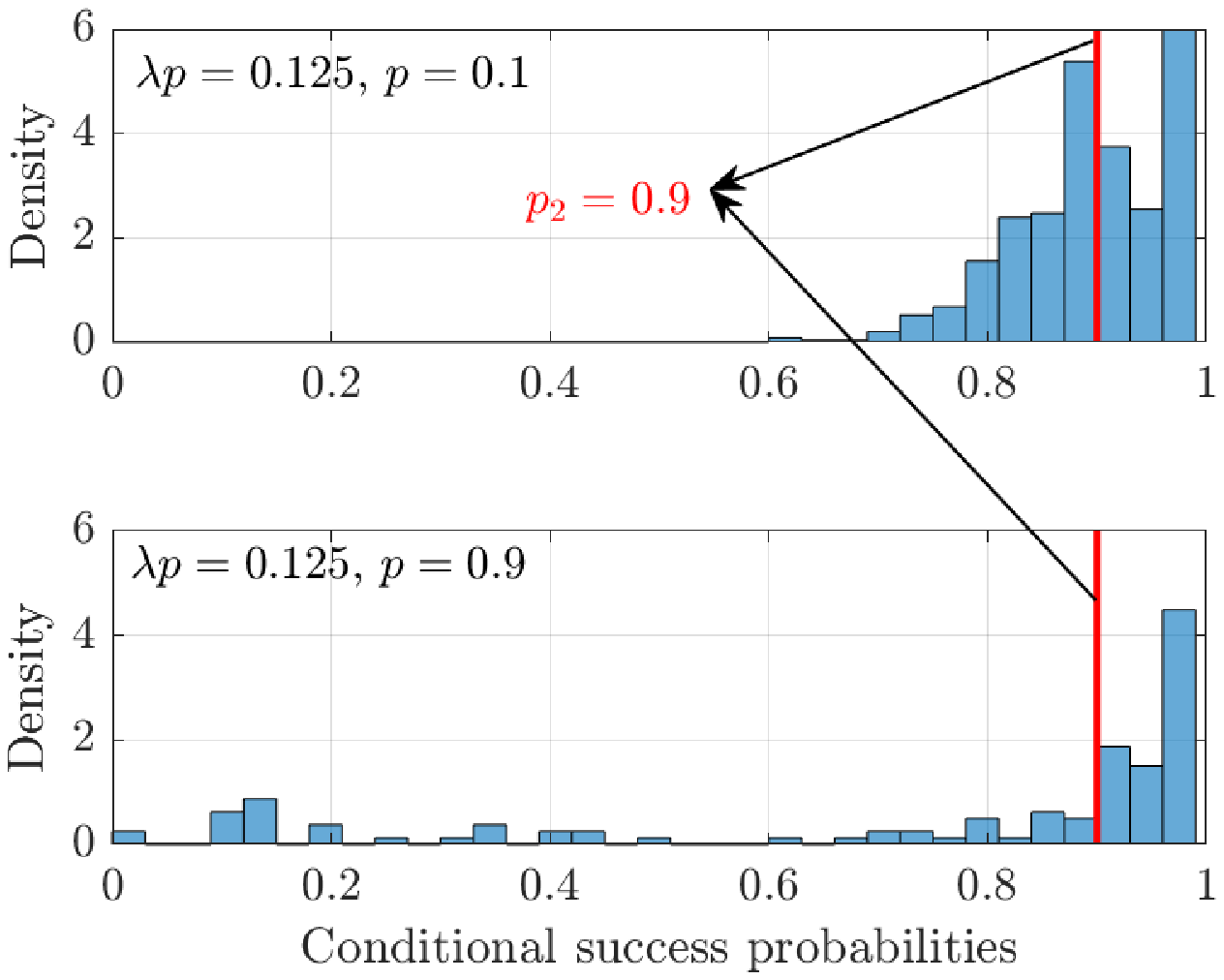}{\label{fig:mean_hist}}}
\caption{\label{fig:mean_lamp} (a) Pairs of ($1/\lambda, p$) such that $p_{2}(\lambda, p) = q$ for  $q = 0.1, 0.5, 0.8, 0.9$ in the TPPP. $\theta = 0$ dB, $D = 0.25$, $\mu = 1$, and $\alpha = 4$. The equation numbers of $p_{2}$ are given in the parentheses in the legends. (b) Histograms of conditional link success probabilities for different combinations of $(1/\lambda, p)$ that yield $p_{2} = 0.9$ in (a).}
\end{figure*}

Fig.~\ref{fig:mean_cc_a} plots the pairs $(1/\lambda,p)$ that satisfy the target success probability of the typical general vehicle $p_{2}(\lambda,p)=q$ for different values of $q$. It is convenient to plot $1/\lambda$ vs. $p$ rather than $\lambda$ vs. $p$ to illustrate the difference between success probability-based and beta approximation-based congestion control methods. It follows from~\eqref{eq:m1_tppp} that $p$ is a linear function of $1/\lambda$, and each line follows a equation of the form $\lambda p = C$, where $C$ is a constant. For a given target $p_{2}$, as $\lambda$ scales by $a$, $p$ is scaled by $1/a$. Further, we observe that for the same $\lambda$, we have to more aggressively reduce $p$ at higher target $p_{2}$ than in the lower target values. 

Fig.~\ref{fig:mean_hist} shows the histograms of conditional success probabilities for different combinations of $(1/\lambda, p)$ picked from the line corresponding to $p_{2} = 0.9$ in Fig.~\ref{fig:mean_cc_a}. We see that for a given $\lambda p$, the conditional success probabilities exhibit higher variance for $p = 0.9$ than for $p = 0.1$. This implies that the fraction of links that are reliable with a probability of at least $x$ varies for different $(1/\lambda, p)$ even though they yield the same success probability. Therefore, to maintain certain link-level reliability, we need to use the SIR MD for congestion control. 

\begin{figure*}[!t]
\centering
\subfloat[$x = 0.1$]{\includegraphics[scale = 0.4]{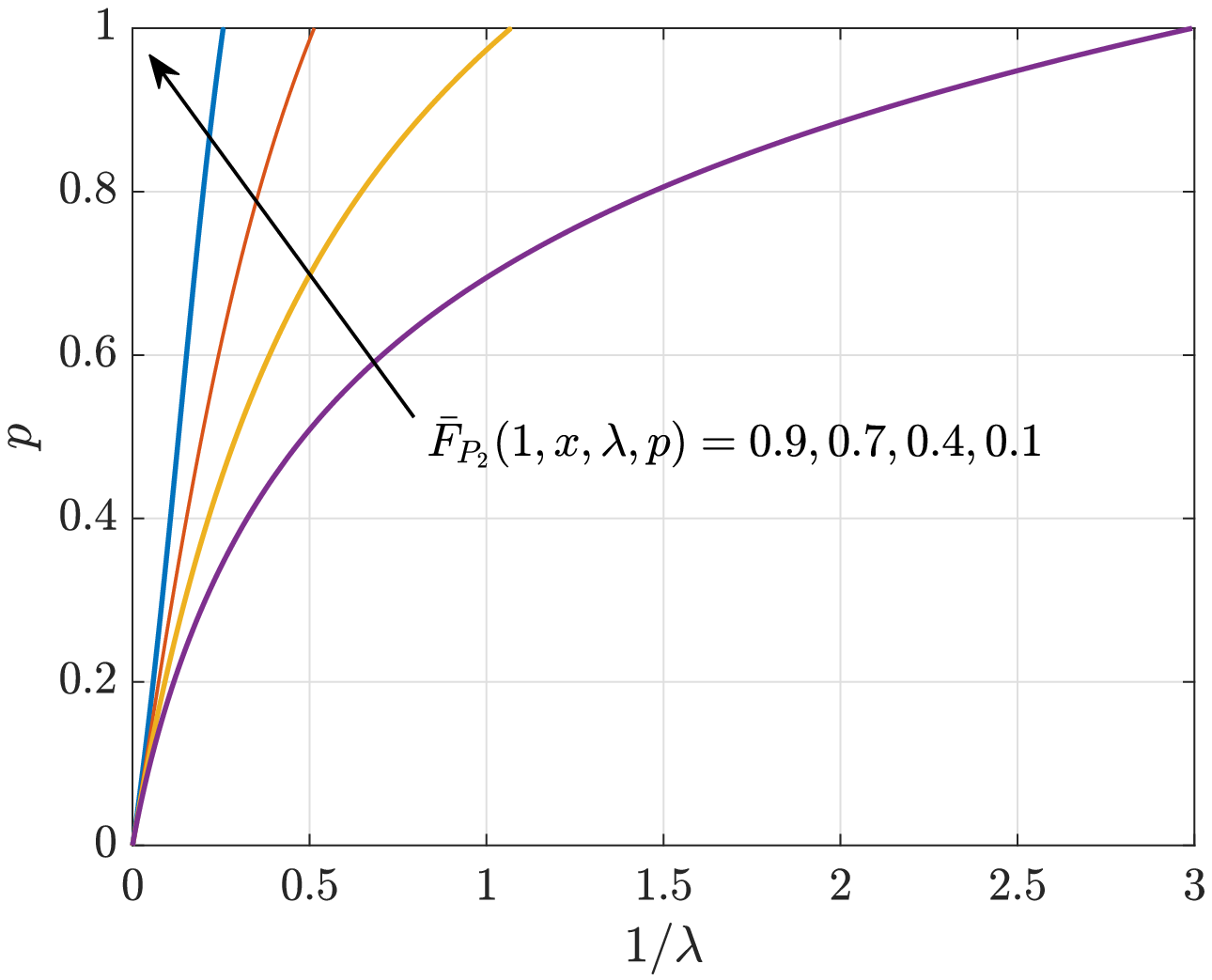}{\label{fig:beta_cc_a}}} \hfill
\subfloat[$x = 0.5$]{\includegraphics[scale = 0.4]{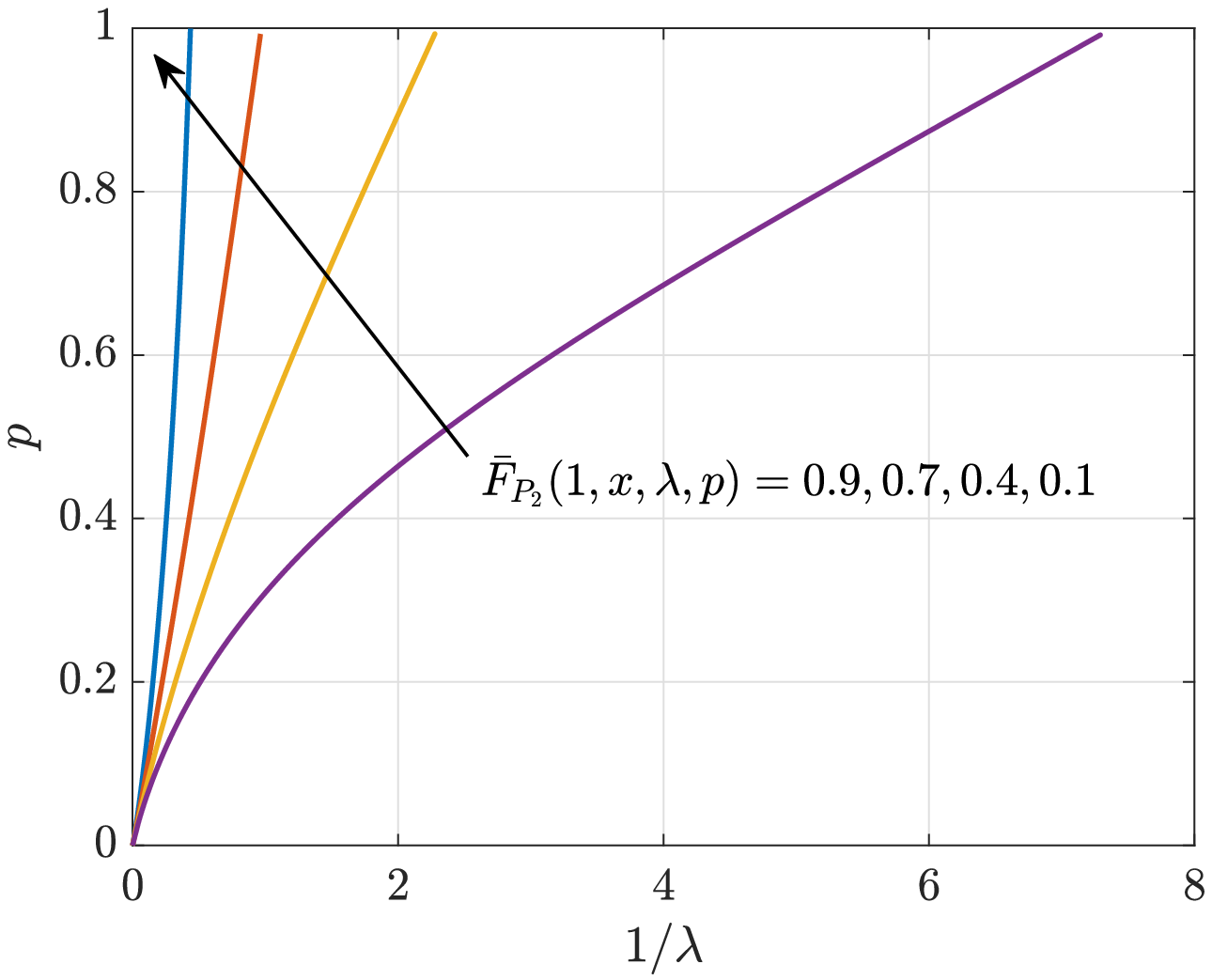}{\label{fig:beta_cc_b}} }\hfill
\subfloat[$x = 0.9$]{\includegraphics[scale = 0.4]{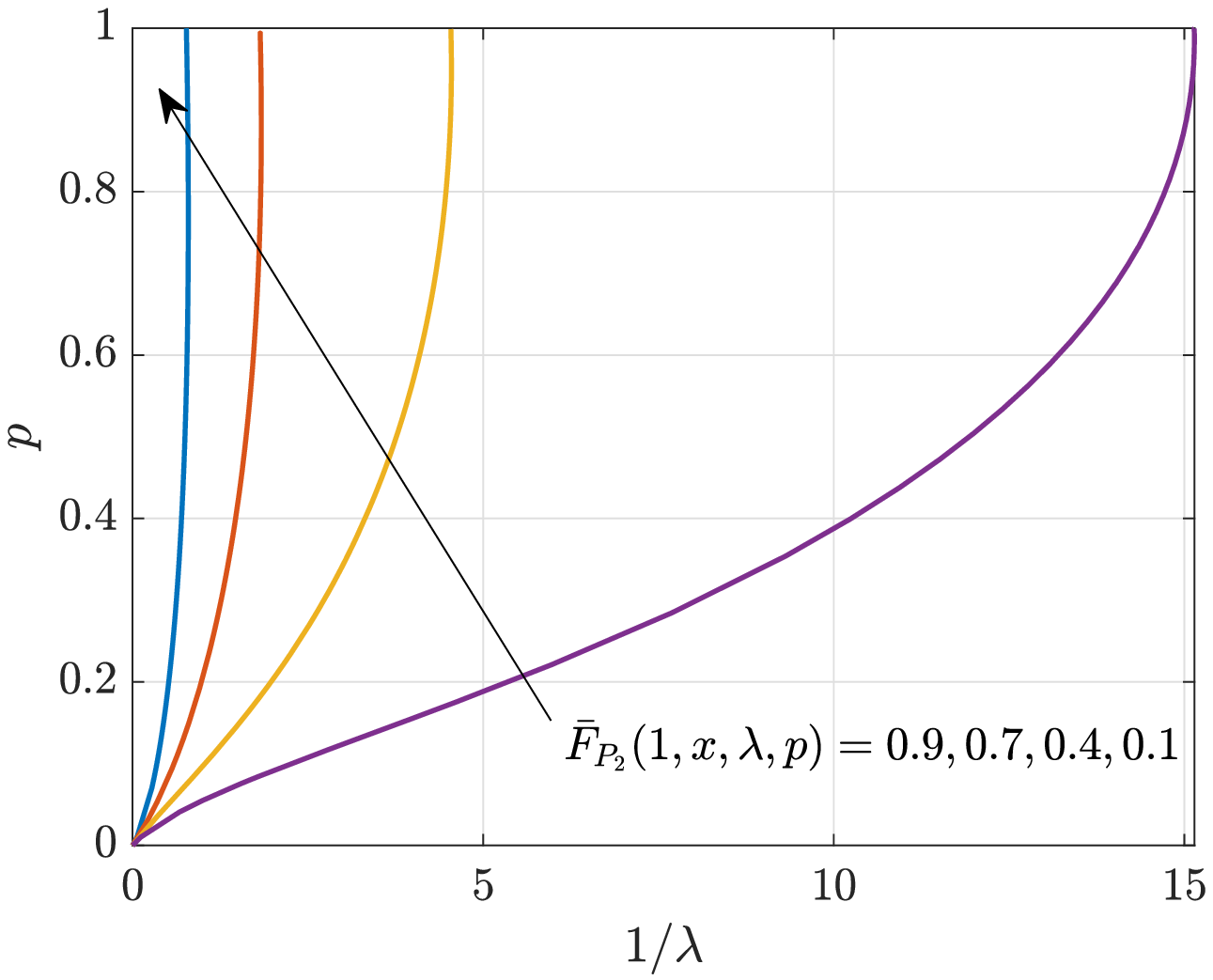}{\label{fig:beta_cc_c}} }
\caption{\label{fig:beta_lamp} Pairs $(1/\lambda,p)$ such that $\bar{F}_{P_2}(1, x,\lambda,p)=q$ for $q=0.1,0.4, 0.7$, and $0.9$. $D = 0.25$, $\mu = 1$, and $\alpha = 4$.  }
\end{figure*}

\begin{figure}
\centering
{\includegraphics[scale = 0.62]{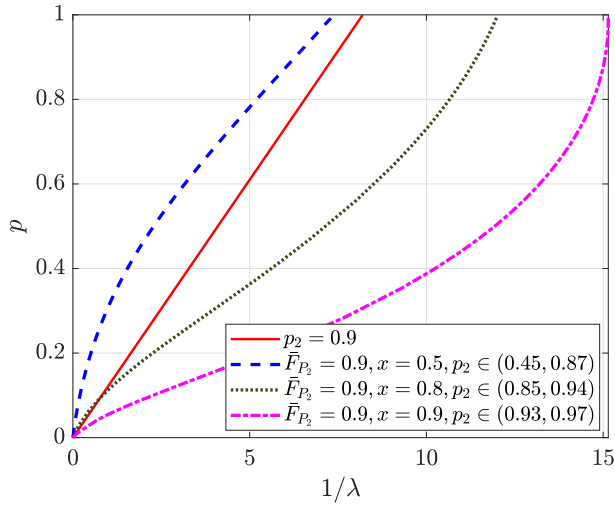}}
\caption{\label{fig:mean_vs_beta} Pairs of ($1/\lambda, p$) that satisfy the target performance given in the legends. The range of $p_{2}$ listed for each combination $(\bar{F}_{P_2}, x)$ is obtained by finding the success probabilities for different values of $(1/\lambda, p)$ sampled along the contour that satisfies $\bar{F}_{P_2}(1, x, \lambda, p) = 0.9$ for a given $x$. $D = 0.25$, $\mu = 1$, and $\alpha = 4$.}
\end{figure}

\subsection{Beta Approximation-Based Congestion Control}
To make the dependence of the MD on $\lambda$ and $p$ explicit, we are adding these two parameters as arguments to the MD as $\bar{F}_{P_2}(\theta, x,\lambda,p)$. 
Fig.~\ref{fig:beta_lamp} plots the pairs $(1/\lambda,p)$ such that $\bar{F}_{P_2}(1, x,\lambda,p)=q$ for different values of $q$ and $x$. We observe that the $(1/\lambda, p)$ contours transition from concave to linear to convex as we increase $x$ for all values of $\bar{F}_{P_2}$. For example, as $\lambda \to \infty$ $(1/\lambda \to 0)$, we can more aggressively vary $p$ at lower values of $x$ than at higher values of $x$. The converse is observed as $\lambda \to 0$ $(1/\lambda \to \infty)$. This implies that we have to change $p$ differently with respect to the reliability constraint rather than simply changing $p$ such that $\lambda p = C$ as in the success probability-based congestion control. In other words, to maintain a certain target fraction of reliable links, $\lambda p$, and, in turn, the success probability, cannot be kept constant. 

Fig.~\ref{fig:mean_vs_beta} presents a different cross-section of Fig.~\ref{fig:beta_lamp} that helps us compare the success probability-based and beta approximation-based congestion control schemes. We observe that the transmit probability $p$ that achieves $p_2=0.9$ (solid line) is higher than the $p$ that guarantees $90 \%$ of links to be at least $80 \%$ reliable (dotted curve), especially, at lower $\lambda$. To guarantee a minimum of $80 \%$ reliability, we shall vary $p$ based on the dotted curve rather than the solid line that yields $p_{2} = 0.9$. This would lower the success probability to as low as $0.85$, implying that we can sacrifice the success probability $p_{2}$ to maintain a certain reliability at each link. Therefore, the success probability is not an adequate measure of congestion when the conditional success probabilities exhibit significant variance, {\em{i.e.,}} when vehicles form a non-regular point process. In contrast, if vehicles form a lattice, the conditional success probabilities are concentrated around the success probability, because the distances to the interferers are the same at each receiver.  

Next, we see whether the inference obtained for congestion control using the beta approximation-based scheme holds for the PLP-PPP. In Fig.~\ref{fig:cc_plp_vs_tppp}, we plot the exact SIR MDs for the PLP-PPP for different values of $\lambda$, with the corresponding $p$ values chosen according to the beta approximation-based congestion control scheme. These $(\lambda, p)$ pairs yield an SIR MD of $0.9$ for the TPPP for $x = 0.5$ and $0.9$ (Figs.~\ref{fig:beta_cc_b} and \ref{fig:beta_cc_c}). We observe in Fig.~\ref{fig:cc_plp_vs_tppp} that the SIR MDs for the PLP-PPP are highly concentrated around $0.9$ with small deviations. This validates that the beta approximation of the SIR MD for the TPPP is sufficient for congestion control. In fact, using the exact SIR MD of the PLP-PPP for congestion control would be prohibitively complicated due to the infinitely many moments involved and their unwieldy expression~\eqref{eq:mb_plp}.

\begin{figure}[t]
\centering
\includegraphics[scale=.62]{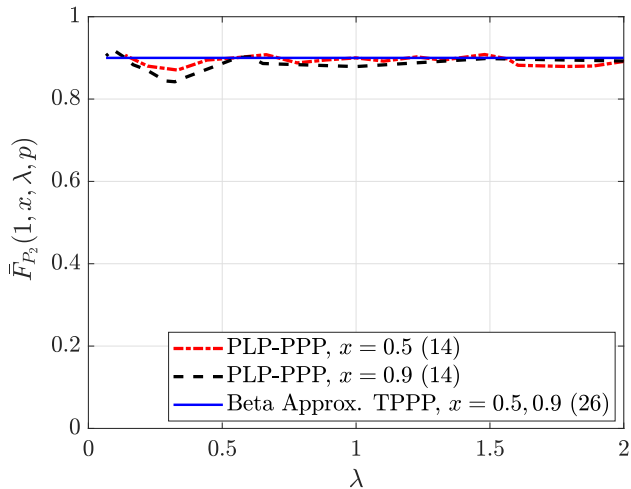}
\caption{\label{fig:cc_plp_vs_tppp}{Comparison of exact SIR MD for the PLP-PPP and beta-approximated SIR MD for the TPPP. For each $\lambda$, $p$ is chosen according to the beta approximation-based congestion control scheme such that the SIR MD for the TPPP equals $0.9$. $\mu = 1$, $D = 0.25$, and $\alpha = 4$. The equation numbers in the parentheses in the legends either refer to the moments or expression of $\bar{F}_{P_{2}}$.}}
\label{fig:cc_plp_vs_tppp}
\end{figure}

\section{Conclusions}
We introduced a simple transdimensional approach to analyze complicated vehicular network models such as the PLP-PPP or PSP-PPP, where the streets are characterized by the PLP or PSP and vehicles on each street form a 1D PPP. The TPPP accounts for only the geometry of the vehicle locations on the street(s) passing through the typical vehicle and models the rest of the vehicles as random points on the 2D plane ignoring their street geometry. Such a transdimensional approach leads to a much simpler and more tractable analysis of the PLP-PPP and PSP-PPP with good accuracy. We showed that under no shadowing, the SIR meta distribution for the TPPP well approximates that for the PLP-PPP and PSP-PPP, and particularly, the approximations are tight in the asymptotic regimes of data rate and reliability. We proved that the SIR meta distribution for the TPPP becomes exact as the variance of the shadowing increases. Hence, it is not essential to account for the geometry of every single street for vehicular network analysis. From the perspective of network simulation, the TPPP model enables us to focus on simpler simulation setups, thus saving computational costs and time.

 We conjecture that the TPPP is sufficient even if the streets are curved segments or circles, etc., rather than lines as in the PLP or sticks as in the PSP. The reason is that any PPP-based vehicular network interloops both the properties of both 1D and 2D PPPs, which indeed is the fundamental principle behind the construction of the TPPP. 

Further, the SIR meta distribution enables network congestion control while ensuring fairness among the links, by guiding the choice of the transmit rate such that each transmitter-receiver link is reliable with a probability of at least $x$. We showed that the success probability or packet reception rate, a measure of the average reliability of the links, is inadequate to understand and alleviate congestion in a network with irregular vehicle spacing since it cannot guarantee that a certain fraction of links achieves the required reliability.  

\appendix
\subsection{Proof of Theorem~\ref{md_plp_result}}
\label{appendix:md_plp_ppp}
By~\eqref{eq:sir_eqn} and \eqref{eq:link_ps}, the conditional success probability can be expressed as
\begin{align}
P^{\mathrm{PLP-PPP}}_{m}(\theta) & = \mathbb{P}(g > \theta D^{\alpha} I \mid I, \mathcal{V}) \nonumber \\
& = \mathbb{E}_{I}[\exp(-\theta D^{\alpha} I) \mid \mathcal{V}] \nonumber \\
& \stackrel{(a)} = \prod _{z \in \mathcal{V}} \bigg( \frac{p}{1+s \Vert z \Vert^{-\alpha}} + 1-p  \bigg)^{b},
\end{align}
where $s = \theta D^{\alpha}$, and $(a)$ is obtained by averaging over ALOHA and fading. Then 
\begin{align}
M^{\mathrm{PLP-PPP}}_{b,m} &=  \mathbb{E}[P_{m}(\theta)^{b}] \nonumber \\
& = \mathbb{E}\bigg[\prod _{z \in \mathcal{V}} \bigg( \frac{p}{1+s \Vert z \Vert^{-\alpha}} + 1-p  \bigg)^{b} \bigg] \nonumber \\
& \stackrel{(b)} =   \underbrace{\mathbb{E}\bigg[\prod _{z \in \mathcal{V}_o^{m} } \bigg( \frac{p}{1+s \Vert z \Vert^{-\alpha}} + 1-p  \bigg)^{b} \bigg]}_{M^{o}_{b,m}} \underbrace{ \mathbb{E}\bigg[\prod _{z \in \mathcal{V}_{!} } \bigg( \frac{p}{1+s \Vert z \Vert^{-\alpha}} + 1-p  \bigg)^{b} \bigg]}_{M^{!}_{b, m}}, \label{eq:mb_prod}
\end{align}
where $(b)$ follows from the independence of the 1D PPPs. $M^{o}_{b, m}$ and $M^{!}_{b, m}$ are the $b$-th moments with respect to the point processes $\mathcal{V}_{o}^{m}$ and $\mathcal{V}_{!}$. We have
\vspace{-2mm}
\begin{align}
M^{o}_{b, m} & \stackrel{(c)}= \exp \bigg(-2 \lambda \int _{0}^{\infty} \bigg[ 1- \bigg( \frac{p}{1+ s u^{-\alpha}} + 1-p \bigg)^{b} \bigg] \mathrm{d}u \bigg) \nonumber \\
& = \exp \bigg(-2 \lambda \int _{0}^{\infty} \bigg[ 1- \bigg( 1-\frac{p s}{u^{\alpha} + s } \bigg)^{b} \bigg] \mathrm{d}u \bigg) \nonumber \\
& \stackrel{(d)} = \exp \bigg(- \lambda \delta \int _{0}^{\infty} \bigg[ 1- \bigg(1- \frac{p s}{v + s } \bigg)^{b} \bigg]v^{\delta/2 -1} \hspace{0.1mm} \mathrm{d}v \bigg),
\end{align}
where $(c)$ applies $\Vert z \Vert = \vert (- u\sin\varphi, u \cos\varphi) \vert_{2}$, and the probability generating functional (PGFL) of the PPP and $(d)$ is the result of substituting $v = u^{\alpha}$. Similarly, we can derive $M^{!}_{b, m}$ as
\begin{align}
 M^{!}_{b, m} & = \mathbb{E}\bigg[\prod _{z \in \mathcal{V}_!} \bigg( \frac{p}{1+s \Vert z \Vert^{-\alpha}} + 1-p  \bigg)^{b} \bigg] \nonumber \\
& \stackrel{(e)} = \mathbb{E}\bigg[\exp \bigg(-\lambda \int _{\mathbb{R}} \bigg[ 1- \bigg( \frac{p}{1+(t^2+u^2)^{-\alpha}} + 1-p \bigg)^{b} \bigg] \hspace{0.1mm} \mathrm{d}u \bigg)\bigg] \nonumber \\
& \stackrel{(f)} = \exp \bigg(-2  \mu \int _{0}^{\infty} ( 1-  G_{b}(t) ) \hspace{0.1mm} \mathrm{d}t  \bigg), \label{eq:Mb1}
\end{align}
where $\Vert z \Vert = \Vert (t\cos \varphi - u\sin\varphi, t\sin\varphi+ u \cos\varphi) \Vert_{2}$ in $(e)$, $(f)$ follows from the PGFL of the PPP, and
\begin{align}
 G_{b}(t) & = \exp \bigg(-2 \lambda \int _{0}^{\infty} \bigg[ 1- \bigg( \frac{p}{1+s (t^{2} + u^{2})^{-\alpha}} + 1-p \bigg)^{b} \bigg] \mathrm{d}u \bigg) \nonumber \\
& = \exp \bigg(-2 \lambda \int _{0}^{\infty} \bigg[ 1- \bigg(1- \frac{p s}{(t^{2}+u^{2})^{1/\delta} + s } \bigg)^{b} \bigg] \mathrm{d}u \bigg) \nonumber \\
& \stackrel{(g)} = \exp \bigg(- \lambda \delta \int _{t^{2/\delta	}}^{\infty} \bigg[ 1- \bigg(1- \frac{p s}{v + s } \bigg)^{b} \bigg]\frac{v^{\delta -1}}{\sqrt{v^{\delta}-t^{2}}} \hspace{0.1mm} \mathrm{d}v \bigg) \label{eq:Gb_prev} \\
& = \exp(-\lambda \delta F_{b}(t)), \label{eq:Gb}
\end{align}
where $(g)$ follows from the change of the variable $v^{2} = t^{2} + u^{2}$. Note that $G_{b}(0) =M^{o}_{b,m}$. Using the binomial expansion, $F_{b}(t)$ can be expanded as~\cite{md_main}
\begin{align}
F_{b}(t) &=   \sum_{k = 1}^{\infty} \binom{b}{k} (p s)^{k}(-1)^{k+1} \int _{t^{2/\delta	}}^{\infty} \frac{v^{\delta -1}}{(v + s )^{k}\sqrt{v^{\delta}-t^{2}}} \hspace{0.1mm}\mathrm{d}v. \label{eq:Fb}
\end{align}
For $t = 0$, \eqref{eq:Gb} reduces to
\begin{align}
G_{b}(0) &= \exp\bigg(-2 \lambda \theta^{\delta/2} D \frac{\pi \delta/2}{\sin(\pi \delta/2)} \sum_{k=1}^{\infty} \binom{b}{k} \binom{\delta/2-1}{k-1}p^{k} \bigg) \nonumber \\
&= \exp\bigg(-2 \lambda D\theta^{\delta/2}  \Gamma(1+\delta/2) \Gamma(1-\delta/2) \mathfrak{D}_{b}(p, \delta/2) \bigg), \label{eq:Gb_exp}
\end{align}
where $\mathfrak{D}_{b}(p, \delta/2) = \sum_{k=1}^{\infty} \binom{b}{k} \binom{\delta/2-1}{k-1}p^{k} = pb \hspace{0.7mm} _{2}F_{1} (1-b, 1-\delta/2; 2; p)$. Substituting \eqref{eq:Gb_exp} for $M_{b,m}^{o}$ and \eqref{eq:Mb1} in \eqref{eq:mb_prod}, we obtain the result in Theorem~\ref{md_plp_result}.

\subsection{Proof of Corollary~\ref{cor_var0}}
\label{appendix:var0}
Corollary~\ref{cor_var0} states that the variance of the conditional success probabilities tends to zero as $p \to 0$ while $\lambda p$ is set to a constant $C$. By~\eqref{eq:mb_prod}, the moment $M_{b,m}$ can be expressed as $
M_{b,m} = M^{o}_{b,m} M^{!}_{b,m}$. 
The first term inside the exponential function in~\eqref{eq:mb_plp} refers to $M^{o}_{b,m}$ and the second term is $ M^{!}_{b,m}$. $\mathfrak{D}_{b}(p, \delta/2) = p$ for $b = 1$ and $2 p + (\delta/2 -1)p^{2}$ for $b = 2$. Thus $M^{o}_{2,m} = (M^{o}_{1,m})^{2 + (\delta/2 -1)p}$. By~\eqref{eq:Mb1}, $M^{!}_{2,m} = \exp (-2  \mu \int _{0}^{\infty} ( 1-  G_{2}(t) \hspace{0.1mm} \mathrm{d}t)$, where 
\begin{align}
G_{2}(t) = \exp \bigg(- \lambda \delta \int _{t^{2/\delta	}}^{\infty} \bigg[ 1- \bigg(1- \frac{p s}{v + s } \bigg)^{2} \bigg]\frac{v^{\delta -1}}{\sqrt{v^{\delta}-t^{2}}} \mathrm{d}v \bigg).
\end{align}

As $p \to 0$ with $\lambda p = C$, we have
\begin{align}
\lim_{\substack{p \to 0 \\ \lambda p = C}} M^{!}_{2,m} &  = \exp \bigg( -2  \mu \lim_{\substack{p \to 0 \\ \lambda p = C}} \int _{0}^{\infty} ( 1-  G_{2}(t)) \hspace{0.1mm} \mathrm{d}t \bigg) \nonumber \\
& \stackrel{(a)} = \exp \bigg( -4  \mu \lambda \delta p s \int _{t^{2/\delta}}^{\infty} \frac{v^{\delta -1}}{(v + s)\sqrt{v^{\delta}-t^{2}}} \hspace{0.1mm} \mathrm{d}v + o(p^{2}) \bigg) \nonumber \\
&\stackrel{(b)}\approx \exp \bigg( -4  \mu \lim_{\substack{p \to 0 \\ \lambda p = C}} \int _{0}^{\infty} ( 1-  G_{1}(t)) \hspace{0.1mm} \mathrm{d}t \bigg)  = (M^{!}_{1,m})^{2}, \label{eq:m2}
\end{align} 
where $(a)$ applies Taylor's series and $(b)$ follows from~\eqref{eq:Gb_prev}. Now, we are ready to evaluate the variance $M_{2,m} -M^{2}_{1,m}$ of the conditional success probability.
\begin{align}
M_{2,m} -M^{2}_{1,m} & = (M^{o}_{1,m})^{2 + (\delta/2 -1)p} M^{!}_{2,m} - (M^{o}_{1,m} M^{!}_{1,m})^{2} \nonumber \\
& = (M^{o}_{1,m})^{2}((M^{o}_{1,m})^{(\delta/2 -1)p}M^{!}_{2,m} - (M^{!}_{1,m})^{2}). \label{eq:m2_lim}
\end{align}
By~\eqref{eq:m2}, as  $p \to 0$ with $\lambda p = C$, \eqref{eq:m2_lim} reduces to
\begin{align*}
 \lim_{\substack{p \to 0 \\ \lambda p = C}}(M^{o}_{1,m} M^{!}_{1,m})^{2} ((M^{o}_{1,m})^{(\delta/2 -1)p} - 1) = 0.
\end{align*}

\bibliographystyle{IEEEtran}
\bibliography{paper}

\end{document}